\documentclass[useAMS,usenatbib,twocolumn]{mn2e}
\usepackage{natbib,amsbsy}
\usepackage{graphicx}
\usepackage{times}
\usepackage{rotate}
\usepackage{color,url}
\newcommand{\simgt}{\lower.5ex\hbox{$\; \buildrel > \over \sim \;$}}
\newcommand{\simlt}{\lower.5ex\hbox{$\; \buildrel < \over \sim \;$}}
\newcommand{\ave}[1]{\left\langle #1\right\rangle}

\newcommand*{\Bell}{\ensuremath{\boldsymbol\ell}}
\newcommand*{\Bgamma}{\ensuremath{\boldsymbol\gamma}}

\begin{document}
\title[Pseudo-spectrum analysis of g-g lensing]{A pseudo-spectrum analysis of galaxy-galaxy lensing}

\author[C. Hikage and M. Oguri]{Chiaki Hikage$^{1}$ and Masamune Oguri$^{2,3,1}$ \\
$^1$ Kavli Institute for the Physics and Mathematics of the Universe (Kavli IPMU, WPI), The University of Tokyo, Kashiwa, Chiba 277-8583, Japan\\
$^2$ Research Center for the Early Universe, University of Tokyo, 7-3-1 Hongo, Bunkyo-ku, Tokyo 113-0033, Japan\\
$^3$ Department of Physics, University of Tokyo, 7-3-1 Hongo, Bunkyo-ku, Tokyo 113-0033, Japan
}
\maketitle

\label{firstpage}

\begin{abstract}
We present the application of the pseudo-spectrum method to 
galaxy-galaxy lensing.  We derive explicit expressions for the
pseudo-spectrum analysis of the galaxy-shear cross spectrum, which is
the Fourier space counterpart of the stacked galaxy-galaxy lensing
profile. The pseudo-spectrum method corrects observational issues such
as the survey geometry, masks of bright stars and their spikes, and
inhomogeneous noise, which distort the spectrum and also mix the
E-mode and the B-mode signals.  Using ray-tracing simulations in
$N$-body simulations including realistic masks, we confirm that the
pseudo-spectrum method successfully recovers the input galaxy-shear
cross spectrum. We also investigate the covariance of the galaxy-shear
cross spectrum using the ray-tracing simulations to show that there is
an excess covariance relative to the Gaussian covariance at small
scales ($k\ga 1h$/Mpc) where the shot noise is dominated in the
Gaussian approximation. We find that the excess of the covariance is
consistent with the expectation from the halo sample variance (HSV),
which originates from the matter fluctuations at scales larger than
the survey area.  We apply the pseudo-spectrum method to the
observational data of Canada-France-Hawaii Telescope Lensing survey
(CFHTLenS) shear catalogue and three different spectroscopic samples
of Sloan Digital Sky Survey Luminous Red Galaxy (SDSS LRG), and Baryon
Oscillation Spectroscopic Survey (BOSS) CMASS and LOWZ galaxies. The
galaxy-shear cross spectra are significantly detected at the level of
$7-10\sigma$ using the analytic covariance with the HSV contribution
included. We also confirm that the observed spectra are consistent
with the halo model predictions with the halo occupation distribution
parameters estimated from previous work. This work demonstrates the
viability of galaxy-galaxy lensing analysis in the Fourier space.
\end{abstract}


\section{Introduction}
Growth of large-scale structure probed by weak gravitational lensing
and galaxy clustering provides a key insight into the nature of dark
energy and dark matter.  Galaxy-galaxy lensing, the cross-correlation
between foreground galaxies and background galaxy image distortions,
is a powerful probe of how the matter distributes around
galaxies. Specifically, galaxy-galaxy lensing has been applied to
various galaxy datasets to study the relation between galaxy
properties and their host dark matter properties
\citep[e.g.,][]{Hoekstra05,Mandelbaum06,Leauthaud12,Velander14,Coupon15}.
This relation, when combined with galaxy clustering measurements,
reduces the systematic uncertainty of galaxy biasing and allows us to
derive useful cosmological constraints
\citep[e.g.,][]{Mandelbaum13,Miyatake15,More15}.  These applications
will grow in the near future when high-quality datasets from various
galaxy imaging and spectroscopic surveys are available, such as Subaru
Hyper Suprime-Cam \citep{HSC,Miyazaki15}, Dark Energy Survey
\citep{DES}, Kilo-Degree Survey \citep{KiDS}, Subaru Prime Focus
Spectrograph \citep{PFS}, Dark Energy Spectroscopic Instrument
\citep{DESI}, Large Synoptic Survey Telescope \citep{LSST}, Euclid
\citep{Euclid}, and WFIRST \citep{WFIRST}.

Two-point statistics including galaxy-galaxy lensing can be studied in
real and Fourier spaces. The power spectrum, defined as the square of
the amplitude of the fluctuation as a function of scale in the Fourier
space, is a fundamental statistics to study the physics in the
evolution of cosmic density fluctuation. The power spectrum has been
playing a central role in the analysis of cosmic microwave background
(CMB) temperature/polarization fluctuation and galaxy clustering. One
of the difficulties in measuring the Fourier-space statistics is the
convolution of various observational effects including survey geometry
and masks with the cosmological fluctuations in the Fourier space. In
particular, weak lensing maps are affected by various observational
issues such as complicated masks due to bright stars and their spikes,
inhomogeneous noise due to signal-to-noise of imaging galaxies, and
intrinsic noise depending on the types of imaging galaxies. Due to the
limited sky area and such complicated masks of imaging sky, the
lensing analysis has been mainly conducted using the real-space
statistics such as aperture-mass dispersion, two-point correlation
functions for cosmic shear, and average $\Delta\Sigma$ for
galaxy-galaxy lensing. In partial sky, the information of the power
spectrum is not identical to that of the two-point correlation due to
the mask, which suggests that the complementary analysis using the
power spectrum is important. 

While real space approaches have been common for the cosmic shear
analysis \citep[e.g.,][]{Kilbinger13,Heymans13,Jee13}, Fourier space
approaches are also of growing popularity in cosmic shear
analysis. For instance, cosmic shear analysis using the
pseudo-spectrum method, which has widely been used in CMB analysis
\citep{Hivon02,Kogut03,Brown03,Bunn03,Smith06,Smith07,Grain09,KimNaselsky10}
including the extraction of B-mode polarization signals
\citep{Smith07}, has been proposed \citep{Hikage11,VanderPlas12,Becker16a} 
and was applied to SDSS \citep{Lin12} and the Canada-France-Hawaii
Telescope Lensing survey (CFHTLenS) data \citep{Kitching14}. There is
another method based on likelihood analysis to measure power spectrum
estimation \citep{Seljak98,HuWhite01}, which was also applied to
CFHTLenS data \citep{Kohlinger16}.  We note that recent cosmic shear
analysis from the DES SV data presented results both in real and
Fourier spaces \citep{Becker16b}. 

In contrast, the galaxy-galaxy lensing analysis in the Fourier space has
attracted little attention. This is presumably due to technical
challenges mentioned above, but in fact the power spectrum analysis of
galaxy-galaxy (or cluster-galaxy) lensing has several advantages.  
First, measuring signals from observations in the Fourier space are
usually faster than that in the real space. This is particularly true when
we are  interested in large-scale cross lensing signals (the so-called
two-halo component region) which contains important cosmological
information
\citep[e.g.,][]{Hu04,Jeong09,OguriTakada11,Covone14,Sereno15,Miyatake16,Umetsu16}. 
Second, analytic calculations of signals and covariances in the Fourier
space are easier than in the real space. Third, the covariance matrix
is more diagonal in the Fourier space, and hence easier to handle.

In this paper, we extend the pseudo-spectrum method for cosmic shear
developed in \citet{Hikage11} to galaxy-galaxy lensing. The advantage
of the pseudo-spectrum method is that the computational speed is much
faster than the likelihood analysis, although the former approach
requires careful corrections  of the mask effect. We show that the
pseudo-spectrum method works on the percent-level accuracy using
ray-tracing simulations and halo datasets. We also study the
covariance of galaxy-galaxy lensing power spectrum using the
ray-tracing simulations to show that the so-called halo sample
variance \citep[HSV;][]{TakadaBridle07,Sato09}, and more generally
speaking super sample covariance \citep[SSC;][]{TakadaHu13,LiHuTakada14},
plays an important role at small angular scales.

Furthermore, we apply the pseudo-spectrum method to publicly available 
observational datasets of CFHTLenS shape data
\citep{Heymans12,Erben13} to measure galaxy-galaxy lensing using
various galaxy samples including Sloan Digital Sky Survey Luminous Red
Galaxy \citep[SDSS LRG;][]{Eisenstein01}, and Baryonic Oscillation
Spectroscopic Survey (BOSS) CMASS and LOWZ galaxy samples
\citep{Eisenstein11}. We evaluate the signal-to-noise ratio of our
measurements and compare with the model predictions using the
analytically estimated covariance including Gaussian and HSV terms.

The paper is organized as follows. In Section~\ref{sec:formalism} we
present basic theoretical formulae of the galaxy-galaxy lensing
spectrum. In Section~\ref{sec:pseudo}, we present the formalism of
pseudo-spectrum method applied to the lensing measurements. In
Section~\ref{sec:sim}, we apply the pseudo-spectrum method to
simulated mock samples to test power spectrum reconstruction
method. We also compare the covariance matrix between simulations and
analytic formulae. In Section~\ref{sec:cfhtlens}, we present the
results on the pseudo-spectrum analysis applied to CFHTLenS and SDSS
galaxy datasets. Section~\ref{sec:conclusion} is devoted to the
conclusion. 

\section{Formalism}
\label{sec:formalism}
In this Section, we review the theoretical formalism of the shear
power spectrum and galaxy-galaxy lensing spectrum \citep[see the reviews of
][]{BartelmannSchneider01}. Throughout the paper, the distance is
expressed in comoving unit.
\subsection{Galaxy-galaxy lensing spectrum}
Weak lensing by the large-scale structure probes the convergence field
$\kappa$ that is defined by the projected mass density field $\delta_m$
with the weight $W^\kappa(z)$
\begin{equation}
\kappa(\mathbf{\theta})\equiv \int \frac{dz}{H(z)} W^\kappa(z)\delta_m(\mathbf{\theta};z),
\end{equation}
where $H(z)$ is the Hubble expansion rate at redshift $z$ and the
weight function $W^\kappa(z)$ is defined as
\begin{equation}
\label{eq:w_kappa}
W^\kappa (z)\equiv \bar\rho_m\Sigma_{\rm crit}^{-1}(z),
\end{equation}
with the mean matter density $\bar\rho_m=\rho_{\rm crit}\Omega_m$.
The inverse of the critical surface density $\Sigma_{\rm crit}^{-1}$ is given by
\begin{eqnarray}
\Sigma_{\rm crit}^{-1}(z) & \equiv &
\int_{z} dz_s p(z_s)\Sigma_{\rm crit}^{-1}(z,z_s), \\
\Sigma_{\rm crit}^{-1}(z,z_s) & =& 
\frac{3H_0^2}{2\rho_{\rm crit}}(1+z)\frac{d_A(z)d_A(z,z_s)}{d_A(z_s)},
\end{eqnarray}
where $d_A(z)$ is the comoving angular diameter distance at the
redshift $z$ and $p(z_s)$ is the redshift distribution function of
source galaxies normalized to be unity as $\int dz_s p(z_s)=1$.  

Galaxy-galaxy lensing, the cross-correlation between the number
density field of foreground galaxies and the shear field of background
galaxies, probes the relationship between matter and galaxy
distribution as a function of scale. Using the Limber approximation
\citep{Limber54}, the galaxy-galaxy spectrum is related to the
three-dimensional (3D) galaxy-shear cross spectrum $P^{gm}(k;z)$ as
\begin{eqnarray}
C_\ell^{g\kappa}\equiv \left\langle \tilde\delta_g\left(k=\frac{\ell}{d_A(z)}\right)
\tilde\kappa_\ell^\ast\right\rangle,~~~~~~~~~~~~~~~~~~~~~~~~~~~~~~~~~~~~~~~~~~~~ \nonumber \\
=\int_{z_{\rm min}}^{z_{\rm max}} \frac{dz}{H(z)}
\left[\frac{W^\kappa (z) W^g (z)}{d_A^2(z)}\right]
P^{gm}\left(k=\frac{\ell}{d_A(z)};z\right), 
\end{eqnarray}
where $\tilde\kappa_\ell$ is the two-dimensional (2D) Fourier
transform of $\kappa(\theta)$ and $\tilde\delta_g(k)$ is the Fourier
transform of the projected number density fields of galaxies from
$z_{\rm min}$ to $z_{\rm max}$. The weight function $W^g(z)$ is the
redshift distribution of foreground galaxies normalized to be unity as
$\int_{z_{\rm min}}^{z_{\rm max}} dz W^g(z)=1$. Usually galaxy-galaxy
lensing is measured in the real space using (differential) projected
mass density around foreground galaxies $\Delta\Sigma(R)$ as 
\begin{equation}
\Delta\Sigma (R)\equiv \int\frac{kdk}{2\pi}P^{g\Sigma}(k)J_2(kR),
\end{equation}
where $J_2(x)$ is the second-order Bessel function. The cross spectrum
of the projected mass density field $\Sigma$ with the projected
clustering of foreground galaxies $P^{g\Sigma}(k)$ is related to the
galaxy-matter power spectrum $P^{gm}(k;z)$ as
\begin{eqnarray}
P^{g\Sigma}(k)=\int_{z_{\rm min}}^{z_{\rm max}} \frac{dz}{H(z)} \bar{\rho}_m W^g(z)P^{gm}(k;z).
\end{eqnarray}
Note that the galaxy-$\Sigma$ cross spectrum $P^{g\Sigma}(k)$ is
independent of the source distribution. In the following analysis, we
simply convert from $C_\ell^{g\kappa}$ to $P^{g\Sigma}(k)$ at the mean
redshift $\bar{z}$ as
\begin{equation}
\label{eq:pk_gsig}
P^{g\Sigma}(k)\simeq d_A^2(\bar{z})\Sigma_{\rm crit}(\bar{z})C_{\ell=kd_A(\bar{z})}^{g\kappa},
\end{equation}
where $d_A^2(z)$ term comes from the conversion from multipole $\ell$
to the wavenumber $k$.

\subsection{Halo model approach to galaxy-galaxy lensing}
In the halo model picture, the galaxy-shear cross spectrum is
separated into one-halo and two-halo components
\begin{equation}
P^{g\Sigma}=P^{g\Sigma {\rm (1h)}}+P^{g\Sigma {\rm (2h)}}.
\end{equation}
The one-halo term reflects the projected mass density profile within
the host dark matter halo, and is given as
\begin{eqnarray}
P^{g\Sigma {\rm (1h)}}(k)=\int_{z_{\rm min}}^{z_{\rm max}} \frac{dz}{H(z)} 
\bar\rho_m W^g(z)\int dM \frac{dn}{dM} \frac{M}{\bar\rho_m} \nonumber \\
\times \tilde{u}_{\rm NFW}(k;M,z) 
[\langle N_{\rm cen}\rangle +\langle N_{\rm sat}\rangle \tilde{p}_{\rm sat}(k;M)],
\end{eqnarray}
where $\tilde{u}_{\rm NFW}$ is the Fourier transform of the projected
NFW profile for the halo with mass $M$ \citep{NFW96,WrightBrainerd00}.
In this paper, we employ the mass-concentration relation presented by
\citet{Duffy08}. $\langle N_{\rm cen}\rangle$ and $\langle N_{\rm
  sat}\rangle$ represent the mean numbers of central and satellite
galaxies, respectively, hosted by the halo with mass $M$ based on the
HOD formalism. The additional function $p_{\rm sat}$ represents the
number density profile of satellite galaxies within the host halo
which takes into account the off-centring of galaxies from the halo
centre \citep{OguriTakada11,Hikage13}.  We use the Gaussian
off-centring profile of satellites with the dispersion of the virial
radius
\begin{equation}
\tilde{p}_{\rm sat}(k;M)=\exp\left[-k^2R_{\rm vir}^2(M)/2\right].
\end{equation}
The two-halo term reflects the halo-matter clustering and given as
\begin{eqnarray}
P^{g\Sigma {\rm (2h)}}(k)=\int \frac{dz}{H(z)} \bar\rho_m W^g(z)~~~~~~~~~~~~~~~~~~~~~~~~~~~~~~~~~~ \nonumber \\
\times \int dM\frac{dn}{dM}[\langle N_{\rm cen}\rangle +\langle N_{\rm sat}\rangle \tilde{p}_{\rm sat}(k;M)]P_{\rm hm}(k,z;M).
\end{eqnarray}
Here we simply describe the halo-matter power spectrum $P_{hm}(k,z;M)$
as the linear matter power spectrum $P_{mm}^{\rm lin}(k,z)$
multiplied with the linear galaxy biasing $b(M,z)$.  When comparing
the observations, we use the fitting formula for both halo mass
function $dn/dM$ and bias $b(M,z)$ \citep{Tinker08,Tinker10} where
halo masses are defined as $M_{200}$, the mass enclosed in a sphere
with an average density of 200 times the comoving matter density.

\section{Pseudo-spectrum analysis}
\label{sec:pseudo}
Here we present the formalism for the pseudo-spectrum analysis of
galaxy-galaxy lensing by extending the pseudo-spectrum approach of
cosmic shear in \citet{Hikage11}. We use the flat-sky approximation
for analyzing CFHTLenS data because the curvature effect is negligible
compared to the sample variance and the computational cost is less
expensive than the full-sky calculation. The full-sky formalism is
presented in Appendix~\ref{sec:app}. The shear field 
$\Bgamma(\hat\mathbf{n})$ defined in a reference Cartesian coordinate
system is decomposed into E-mode and B-mode components by the
following Fourier transform
\begin{equation}
\tilde{E}_{\Bell} \pm i\tilde{B}_{\Bell}=\int d\hat\mathbf{n}
\Bgamma(\hat\mathbf{n}) \exp(i\Bell\mathbf{\cdot\hat{n}}\pm 2\varphi_{\Bell}),
\end{equation}
where $\varphi_{\Bell}$ is the azimuthal angle of $\Bell$. Their auto
and cross spectra are defined as
\begin{equation}
\langle X_{\Bell}Y_{\Bell'}^\ast \rangle 
\equiv (2\pi)^2\delta_{\rm D}^2(\Bell-\Bell')C_\ell^{XY},
\end{equation}
where $X$ and $Y$ denotes E- ($\tilde{E}_{\Bell}$)
or B-mode ($\tilde{B}_{\Bell}$). In the weak lensing
field, the E-mode field corresponds to the convergence field and thus
its power spectrum reduces to the convergence power spectrum,
$C_\ell^{\kappa\kappa}=C_\ell^{EE}$. In the standard $\Lambda$ cold
dark matter (CDM) model, the B-mode power and EB cross spectra are
negligibly small and can be used to probe observational 
systematics. The galaxy-shear cross spectrum is also given by 
cross-correlating the projected galaxy distribution with the E-mode
shear as 
\begin{equation}
\ave{\tilde\delta_g\left(\mathbf{k}=\frac{\Bell}{d_A(z)}\right)\tilde{E}_{\Bell'}^\ast}
\equiv (2\pi)^2 \delta_{\rm D}^2(\Bell-\Bell')C_{\Bell}^{gE}.
\end{equation}
The cross-correlation of the galaxy distribution with the B-mode shear
is also negligible and can be used to probe systematics.

An observed imaging field has a finite survey area with complicated
masks. We take account of this mask effect using pseudo-$C_\ell$
method to reconstruct the original shear spectrum deconvolved with the
survey mask. The weak lensing shear is usually estimated from 
observed ellipticities of background galaxy images. The galaxy
ellipticity has a large intrinsic component and also the low
signal-to-noise images are subject to the measurement noise
\begin{equation}
\mathbf{e}^{\rm (obs)}(\mathbf{\hat{n}}) = 
\Bgamma (\mathbf{\hat{n}})+\mathbf{\epsilon_{\rm noi}},
\end{equation}
where $\epsilon_{\rm noi}$ includes the intrinsic ellipticity and
the measurement noise.  Ideally, the shear value is obtained by the
average over observed ellipticities $\Bgamma= \langle
\mathbf{e}^{\rm (obs)} \rangle$.  However, the observed shear map is
masked due to bright stars in our Galaxy and thus the survey mask has
complicated shape due to bright stars and their spikes.  When the grid
is completely inside the mask, we do not obtain any information on the
shear in the grid. Furthermore the expected error of shear in each
grid depends on the number of source galaxies and the uncertainty of
observed ellipticities, which can differ at different positions on the
sky. 

We thus estimate the weight for shear field by summing ellipticity
weights in each pixel
\begin{equation}
U^\gamma(\mathbf{\hat{n}})=\sum_i^{\mathbf{\hat{n}}_i\in\mathbf{\hat{n}}} w_i^\gamma,
\end{equation}
where the shear weight of $i$-th source galaxy $w_i^\gamma$ is
introduced to enhance the signal-to-noise of weak lensing
measurements. In practical analysis, it is common to define the weight
of each galaxy by the inverse variance of the shape noise
$\epsilon_{\rm noi}$. The weight field takes account of the mask
effect by setting the value of $U^\gamma$ to be zero when the grid at
$\mathbf{\hat{n}}$ is completely masked. The observed shear field is
related to the true shear field as
\begin{eqnarray}
\left\langle \Bgamma^{\rm (obs)}(\mathbf{\hat{n}}) \right\rangle
&=&\left\langle\sum_i^{\mathbf{\hat{n}}_i\in\mathbf{\hat{n}}}w_i^\gamma\mathbf{e}_i^{\rm (obs)}\right\rangle,\nonumber \\
&=& U^\gamma(\mathbf{\hat{n}})\Bgamma^{\rm (true)} (\mathbf{\hat{n}}). 
\end{eqnarray}
The weight field of $\Sigma$ is similarly estimated by changing the
weight $w_i^\gamma$ to
\begin{equation}
w_i^\Sigma=\left[\Sigma_{\rm crit}^{-1}(z;z_{s,i})\right]^{2}w_i^\gamma .
\end{equation}

The galaxy number density field is given by the excess of the observed
galaxy number relative to the averaged number density at each
grid. The observed galaxy distribution has different survey mask and
angular selection function. For spectroscopic galaxy samples, there
are more complicated observational effect such as the fiber collision
that suppress the number of close pairs of galaxies. The observed
field is related to the true field as 
\begin{equation}
\left\langle \delta_g^{\rm (obs)}(\mathbf{\hat{n}})\right\rangle
=U^g(\mathbf{\hat{n}})\delta_g^{\rm (true)}(\mathbf{\hat{n}}),
\end{equation}
where $U^g$ is the weight (mask) field of the galaxy number density
field. Here we focus on the small patch of the sky where the flat-sky
approximation holds well. The Fourier transform of the shear field and
the galaxy number density field are affected by the convolution of
the masks
\begin{equation}
\tilde{E}_{\Bell}^{\rm (obs)}\pm i\tilde{B}_{\Bell}^{\rm (obs)}=\frac{1}{\theta_s^2}\sum_{\Bell'}
(\tilde{E}_{\Bell'}^{\rm (true)}\pm i\tilde{B}_{\Bell'}^{\rm (true)})\tilde{U}^\gamma_{\Bell-\Bell'}e^{\pm 2i\varphi_{\Bell'\Bell}},
\end{equation}
where $\varphi_{\Bell'\Bell}\equiv \varphi_{\Bell'}-\varphi_{\Bell}$
and
\begin{equation}
\tilde\delta^{\rm (obs)}_{g,\Bell}
=\frac{1}{\theta_s^2}\sum_{\Bell'}\tilde{U}_{\Bell-\Bell'}^g\tilde{\delta}^{\rm (true)}_{g,\Bell'},
\end{equation}
where we have assumed that the patch is a square field with the side
length of $\theta_s$.

The auto and cross spectra in the sky field are defined as
\begin{eqnarray}
\langle X_{\Bell}Y^{*}_{\Bell'}\rangle&=&\theta_s^2\delta^{\rm K}_{\Bell-\Bell'}C_{\Bell}^{XY},
\end{eqnarray}
where $X$ and $Y$ denote either E-mode shear, B-mode shear, or the
galaxy number density field, and $\delta^{\rm K}_{\Bell}$ is
the Kronecker's delta. Again we do not take into account the imaginary
part of cross spectra. Due to the convolution with the mask (weight)
field in the real space, the power spectrum for the mask (weight)
field has the mode coupling as
\begin{equation}
\mathbf{C}_{\Bell}^{\rm (obs)}=\sum_{\Bell'}\mathbf{M}_{\Bell\Bell'}F_{\ell'}^2C_{\Bell'}^{\rm (true)}+\mathbf{N}_{\Bell}^{\rm (obs)},
\end{equation}
where we introduce the 6-dimensional vector
$\mathbf{C}_{\Bell}=(C_{\Bell}^{EE},C_{\Bell}^{BB},C_{\Bell}^{EB},C_{\Bell}^{gE},C_{\Bell}^{gB},C_{\Bell}^{gg})$,
$\mathbf{M}$ is $6\times 6$ convolution matrix, $F_\ell$ is the pixel
window function and $\mathbf{N}_{\Bell}^{\rm (obs)}$ is the convolved
noise spectrum.  For the lensing power spectrum, the E-mode and B-mode
power spectra are mixed as
\begin{eqnarray}
\left(\begin{array}{c}
\tilde{C}_{\Bell}^{\rm EE (obs)} \\
\tilde{C}_{\Bell}^{\rm BB (obs)} \\
\end{array}\right)
=
\frac{1}{\theta_s^2}\sum_{\Bell'}{\cal U}_{\Bell-\Bell'}^{\gamma\gamma}
\left(\begin{array}{cc}
\cos^2(2\varphi_{\Bell\Bell'}) & \sin^2(2\varphi_{\Bell\Bell'}) \\
\sin^2(2\varphi_{\Bell\Bell'}) & \cos^2(2\varphi_{\Bell\Bell'}) 
\end{array}\right)&& \nonumber \\
&&\hspace{-4cm}\times
\left(\begin{array}{c}
C_{\Bell}^{\rm EE (true)} \\
C_{\Bell}^{\rm BB (true)} \\
\end{array}\right).
\end{eqnarray}
The convolution with the mask generates the B-mode spectrum leaked from
the E-mode spectrum even if there is no intrinsic B-mode power. On the
other hand, galaxy-shear cross spectra are written as
\begin{equation}
C^{gX {\rm (obs)}}_{\Bell}=\frac{1}{\theta_s^2}
\sum_{\Bell'}C^{gX {\rm (true)}}_{\Bell'}{\cal U}_{\Bell-\Bell'}^{g\gamma}\cos(2\varphi_{\Bell\Bell'}),
\end{equation}
where $X$ denotes E- or B-mode shear field. E-mode and B-mode
components do not mix for the galaxy-shear cross spectra because of
their different parity. The EB-mode cross spectrum and
the galaxy auto power spectrum are respectively written as
\begin{equation}
C_{\Bell}^{\rm EB (obs)}=\frac{1}{\theta_s^2}\sum_{\Bell'}C_{\Bell'}^{\rm EB (true)}{\cal U}^{\gamma\gamma}_{\Bell-\Bell'}
[\cos^2(2\varphi_{\Bell\Bell'})-\sin^2(2\varphi_{\Bell\Bell'})],
\end{equation}
and
\begin{equation}
C_{\Bell}^{gg {\rm (obs)}}=\frac{1}{\theta_s^2}\sum_{\Bell'}C_{\Bell'}^{gg {\rm (true)}}{\cal U}^{gg}_{\Bell-\Bell'}.
\end{equation}
In the above expressions, the function ${\cal U}_{\Bell}$ represent
the auto and cross power spectra of shear and galaxy weight fields,
i.e.,
\begin{eqnarray}
\langle \tilde{U}_{X,\Bell}\tilde{U}^*_{Y,\Bell'}\rangle&=&\theta_s^2\delta^{\rm K}_{\Bell-\Bell'}{\cal U}_{\Bell}^{XY},
\end{eqnarray}
with $X$ and $Y$ being $g$ or $\gamma$. Note that the two weight field
$U^\gamma(\mathbf{\hat{n}})$ and $U^g(\mathbf{\hat{n}})$ are not
necessarily identical; our formalism works even if the shear and
galaxy density fields have different mask patterns.  We also
take account of the effect of the finite square field to compute the
full mode coupling matrix $M_\mathbf{\ell\ell'}$ \citep{Hikage11}.

We invert the mode coupling matrix after binning. To do so, we compute
the binned non-dimensional power spectrum  as
\begin{equation}
{\cal C}_b\equiv \frac{1}{N_{{\rm mode},b}}\sum_{\Bell}^{\ell \in \ell_b}P_{b\ell}C_{\Bell},
\end{equation}
where $P_{b\ell}=\ell^2/2\pi$ and $N_{\rm mode,b}$ is the number of
modes in $b$-th bin. The unmasked binned power spectrum is obtained by
multiplying the inverse of the mode coupling matrix with the
pseudo-spectrum 
\begin{equation}
\label{eq:cb}
{\cal C}_b^{\rm (true)}=\mathbf{M}_{bb'}^{-1}\sum_{\Bell}^{\ell\in \ell_b'}P_{b'\ell}
(\mathbf{C}_{\Bell}^{\rm (obs)}-\langle \mathbf{N}_{\Bell}\rangle_{\rm MC}),
\end{equation}
where 
\begin{equation}
\mathbf{M}_{bb'}=\sum_{\ell}^{\ell\in l_b}\sum_{\ell'}^{\ell'\in\ell_{b'}}
P_{b\ell}\mathbf{M}_{\Bell\Bell'}F_{\ell'}^2Q_{\ell'b'},
\end{equation}
with $Q_{\ell b}=2\pi/\ell^2$. In order to remove the shot noise
effect, we randomly rotate ellipticities of individual weak lensing
galaxies to estimate the shot noise power spectrum $\langle
\mathbf{N}_{\Bell}\rangle_{\rm MC}$ and to subtract it from the
power spectrum. In order to obtain accurate estimates of the shot
noise power spectrum, we generate 100 realizations of rotated shear
fields and take their average of $\tilde\mathbf{N}_{\Bell}$. Throughout the
paper we use 15 $\ell$ bins in the range $100\le \ell \le 10800$ that
are equal spaced in the logarithmic scale.

\section{Testing the pseudo-spectrum method using ray-tracing simulations}
\label{sec:sim}

\subsection{Ray-tracing simulations and halo samples}
Ray-tracing simulations in $N$-body simulations have been used to
study the properties of weak lensing fields
\citep[e.g.,][]{Jain00,Hamana01}. We use ray-tracing simulations and
the halo dataset constructed by \citet{Sato09} to check the accuracy
of our pseudo-spectrum method.  We use 400 realizations of shear field
which has the square field with a side length $\theta_s=5$ degree and
the pixel number of $N_{\rm pix}=2048^2$.  The redshift of the source
galaxy is set to $z_s=1$.  At $z<1$, the mass field is obtained from 
$N$-body simulations with $L_{\rm box}=240h^{-1}$Mpc and $256^3$ particles
(each particle mass is $5.44\times 10^{10}h^{-1}M_\odot$) at the
initial redshift of 50. Cosmological parameters in these simulations
are those in the flat $\Lambda$ CDM model based on the {\it WMAP} 3-year
result \citep{Spergel07}: $\Omega_m=0.238$, $\Omega_b=0.042$,
$\Omega_{\Lambda}=0.762$, $\sigma_8=0.76$, $h=0.732$, $n_s=0.958$
(hereafter we denote {\it WMAP3}).

The simulated source galaxies are distributed randomly with the
mean angular number density of $n_{\rm s,gal}=20$ arcmin$^{-2}$. Each
source galaxy takes the shear value at the nearest pixel point in a
given shear field. We add an intrinsic shape noise to each ellipticity
component assuming a Gaussian distribution with the dispersion per
component of $\sigma_{\rm int}=0.22$. We take into account the mask due to
bright stars and their diffraction spikes as described in
\citet{Hikage11}. The radii of the simulated star mask $r$ randomly 
distributes from 0.2 to 2 arcmin.  For each star mask with $r>1$
arcmin, a rectangular shape mask with $0.2r\times 5r$ along $y$-axis
is added to mimic the diffraction spike. We remove source galaxies inside
the masks so that 75\% of the total area is available after masking.  The
shear field with the inverse variance weight becomes
\begin{equation}
\Bgamma'(\mathbf{x})=\sum_i^{\mathbf{x}_{s,i}\in\mathbf{x}}w_i(\Bgamma_i+\epsilon_{\rm int}),
\end{equation}
and the shear weight field is
$U^\gamma(\mathbf{x})=\sum_i^{\mathbf{x}_{s,i}\in\mathbf{x}}w_i$. For
simplicity we set the noise variance for all the simulated source
galaxies to  be same (i.e., $w_i=$const) and thus the shear weight
is simply proportional to the source number density.  Even when $w_i$
is constant, the shear weight field is fluctuated depending on the
number distribution of source galaxies. When a pixel $\mathbf{x}$ is partially
(completely) masked, the relative weight for the pixel becomes less
than unity (zero). 

The ray-tracing simulations also contain haloes identified from the
$N$-body simulations used for ray-tracing. We use the haloes with the
mass $M_{\rm h}\ge 10^{13}h^{-1}M_\odot$ and $M_{\rm h}\ge
10^{14}h^{-1}M_\odot$ at the redshift range of $0.4<z<0.6$ as a
foreground lens sample. The masked density fluctuation of these
foreground ``galaxies'' is obtained by using the data and random as
follows:
\begin{equation}
\delta_g'(\mathbf{x})=n_g(\mathbf{x})-\bar{n}(\mathbf{x}),
\end{equation}
where $\bar{n}(\mathbf{x})$ is the mean number density estimated from
random and and the resulting weight field for the galaxies
$U^g(\mathbf{x})=\bar{n}(\mathbf{x})$. The sky areas covered by
imaging surveys and spectroscopic surveys are usually different. We
mask different 25\% areas in the simulated halo fields and the source
fields. The overlapped area reduces to be 50\% of the original area.

\subsection{Reconstruction of the input spectra}
\begin{figure*}
\begin{center}
\includegraphics[width=8cm]{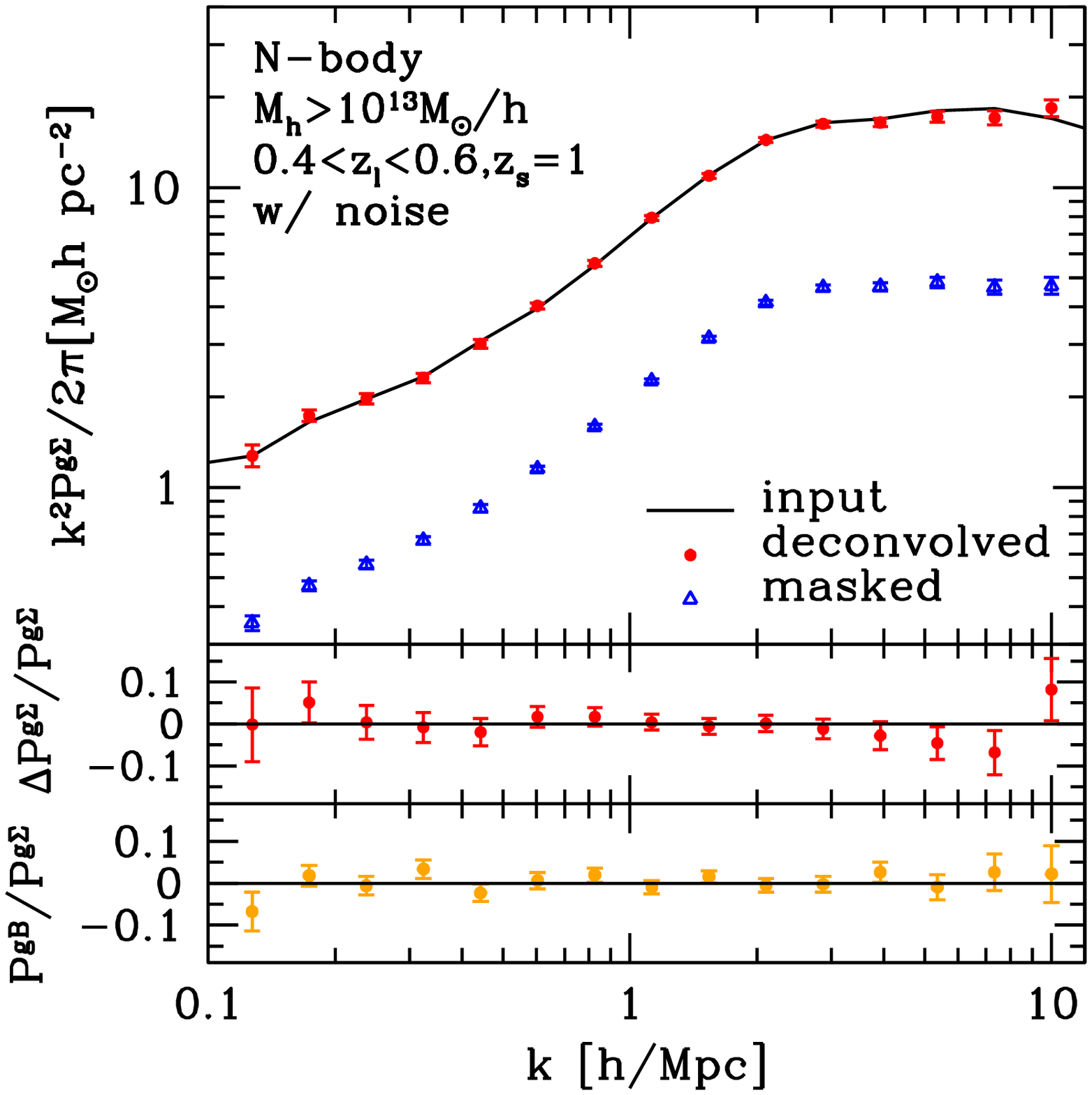}
\includegraphics[width=8cm]{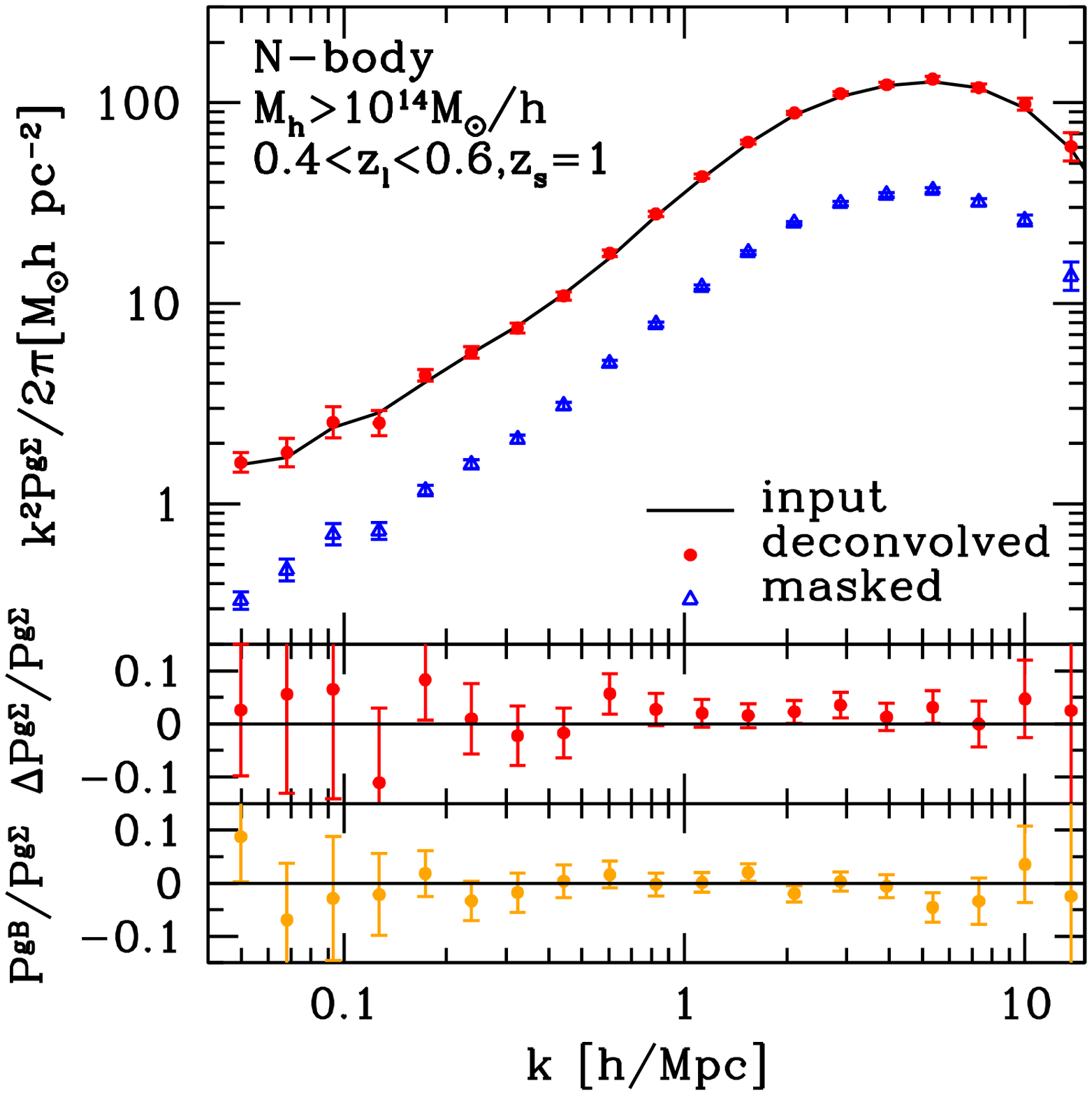}
\caption{Reconstruction of the galaxy-shear cross spectrum
  $P^{g\Sigma}(k)$, i.e., cross-correlation of the weak lensing shear
  field and the halo number density field, using 400 realizations of
  ray-tracing simulations and halo catalogues. Each realization covers
  $25$~deg$^2$. The mask regions are different between the shear and
  halo umber density fields, and the fraction of overlapping area is
  50\% (see text for details). The source redshift is
  $z_s=1$ and the angular number density of source galaxies for weak
  lensing is $20$~arcmin$^{-2}$. The intrinsic shape noise is
  Gaussian with $\sigma_{\rm int}=0.22$. We use haloes with minimum
  halo masses of $10^{13}h^{-1}M_\odot$ ({\it left}) and
  $10^{14}h^{-1}M_\odot$ ({\it right}).
  {\it Top:} The input cross spectra are plotted by solid lines.  The
  deconvolved spectra and masked spectra are respectively plotted by 
  red filled circles and blue open triangles. {\it Middle:} The
  difference ratio between the deconvolved spectrum and the input
  spectrum. {\it Bottom:} The cross-correlation between halo density
  fields and the lensing B-mode field.}
\label{fig:clsig_sim}
\end{center}
\end{figure*}

Fig.~\ref{fig:clsig_sim} shows the results of simulated galaxy
(halo)-shear cross spectra $P^{g\Sigma}(k)$, which is essentially the
galaxy-galaxy lensing profile in the Fourier space. The errorbars
represent the 1$\sigma$ error for the averaged spectra obtained by
computing the dispersion of the spectra reconstructed using the
pseudo-spectrum method divided by the square root of the realization
number, 400. We confirm that the deconvolved spectra (red symbols)
recover the input galaxy-shear cross spectra (black solid lines) for
both cases with the halo mass $M_{h}>10^{13}Mh^{-1}_\odot$ and
$M_h>10^{14}Mh^{-1}_\odot$. The masked spectra (blue symbols) have
different amplitudes and shapes.  As shown in the middle panels of
Fig.~\ref{fig:clsig_sim}, the difference ratios of the deconvolved
spectra to the input spectra are within the errors for a wide range
from 0.1 to 10$h$/Mpc scale. The bottom panels show the cross-spectra
between the halo density fields and the B-mode lensing fields, which
are also consistent with zero. This analysis clearly indicates that
our pseudo-spectrum method accurately recovers the input galaxy-shear
cross spectrum, even in the presence of realistic masks.

\subsection{Covariance of galaxy-galaxy lensing}
\label{sec:cov}
Our extensive galaxy-shear cross-spectrum analysis using ray-tracing
simulations also enables us to quantify the covariance of
galaxy-galaxy lensing. Although the weak lensed field is non-Gaussian 
\citep[e.g.,][]{Takada09,KayoTakadaJain13,SatoNishimichi13}, observed
weak lensed fields are dominated by the shape noise on small scales,
and thereby the Gaussian approximation has often been adopted. In
Gaussian approximation, the covariance matrix of galaxy-shear cross
spectrum is diagonal and is described by
\begin{eqnarray}
\label{eq:err_ga}
{\rm Cov}_{g\Sigma}^{\rm (G)}(k_i,k_j)
=\frac{\delta^{\rm K}_{ij}}{N_{{\rm mode},i}}
\left[P^{g\Sigma~2}(k_i)+\tilde{P}^{gg}(k_i)\tilde{P}^{\Sigma\Sigma}(k_i)\right],
\end{eqnarray}
where the power spectrum $\tilde{P}$ denotes the power spectrum
including shot noise. At small scales, this covariance matrix is
dominated by the shot noise term $\propto \sigma_{\rm
  int}^2/(n_{\rm s,gal}\bar{n}_g)$, where $\bar{n}_g$ is the average
density of foreground galaxies \citep[see, e.g.,][]{OguriTakada11}. The
number of independent $k$-modes in $i$-th bin, $N_{{\rm mode},i}$, is given as
\begin{equation}
N_{{\rm mode},i}=\frac{\pi(k_{i,{\rm max}}^2-k_{i,{\rm min}}^2)\Omega_{\rm sky}^{\rm overlap}}{(2\pi)^2},
\end{equation}
where $\Omega_{\rm sky}^{\rm overlap}$ is the overlapped sky area
between halo maps and shear maps.

\begin{figure*}
\begin{center}
\includegraphics[width=7cm]{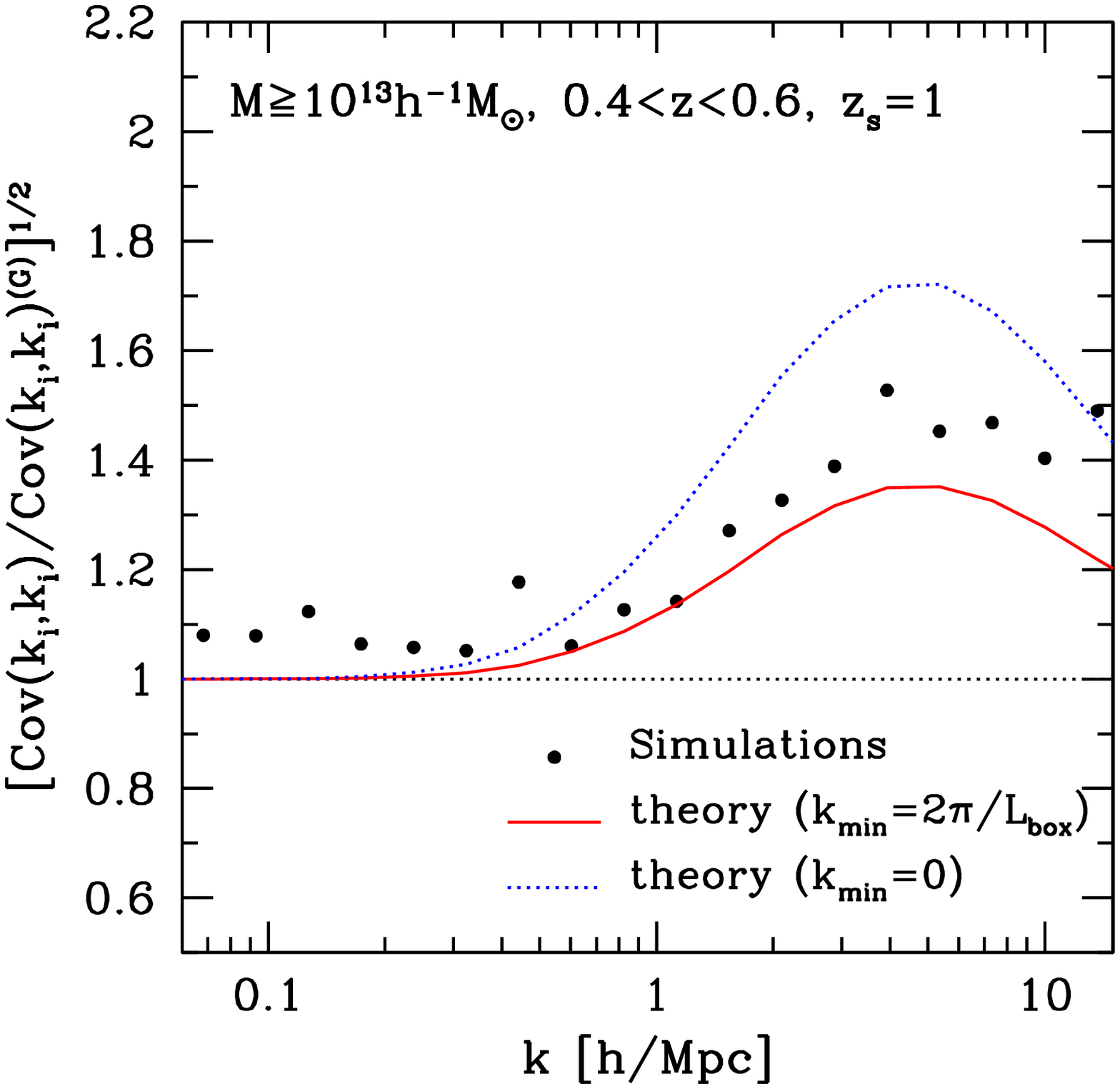}
\includegraphics[width=7cm]{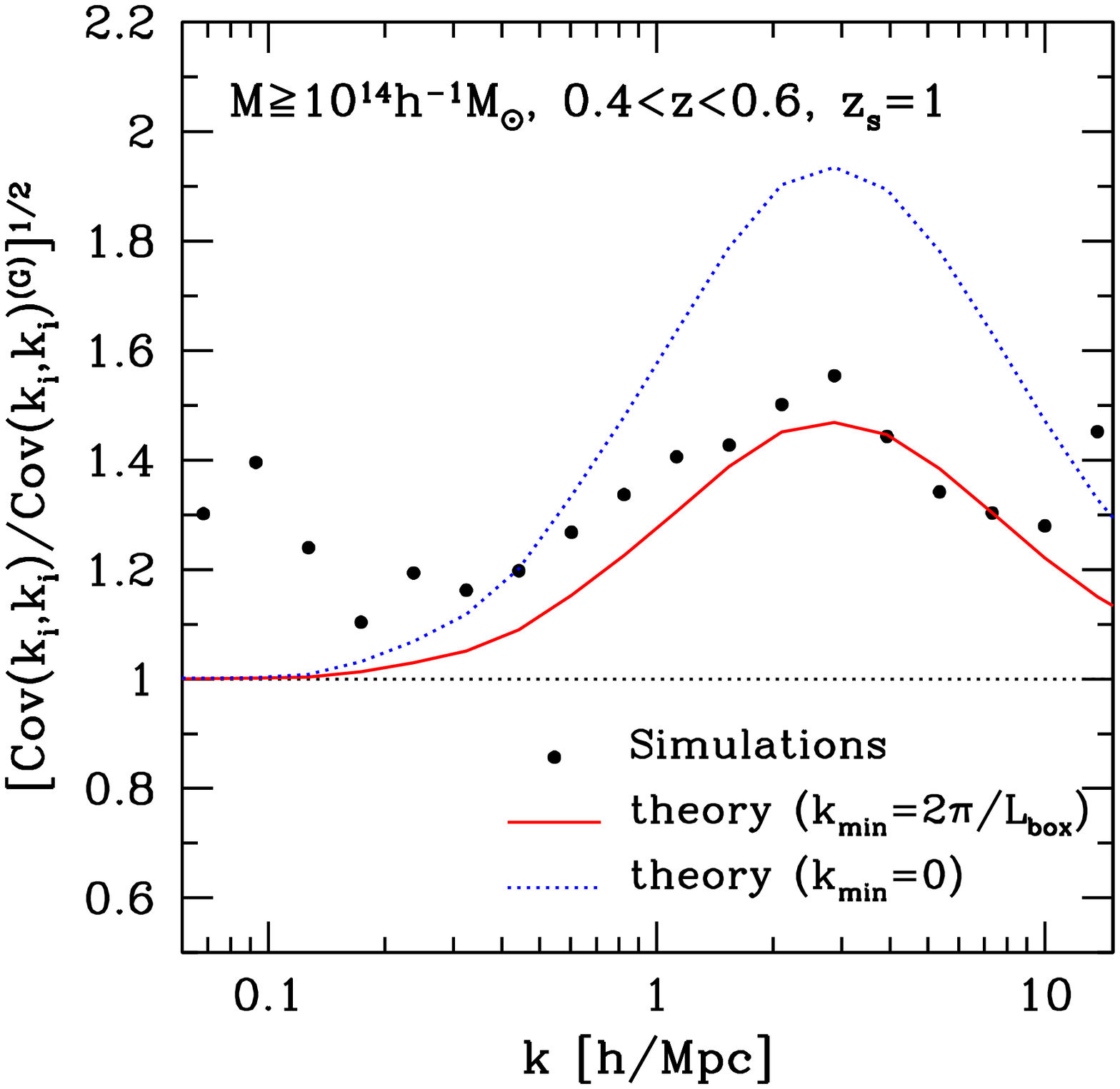}
\caption{Diagonal errors of the simulated deconvolved galaxy-shear
  cross spectra relative to the Gaussian errors
  (eq.~[\ref{eq:err_ga}]). Left and right panels show the results for
  the halo masses of $10^{13}h^{-1}M_\odot$ and
  $10^{14}h^{-1}M_\odot$, respectively. Lines represent the
  theoretical expectations from the HSV including fluctuations up to
  the simulation box size ({\it solid}) and including fluctuations at
  all scales ({\it dotted}).}
\label{fig:cl_sim_error}
\end{center}
\end{figure*}

Fig. \ref{fig:cl_sim_error} compares the errors of diagonal components
for galaxy-shear cross spectrum relative to the Gaussian
expectations. The Gaussian errors for simulations is computed using
the simulated spectrum. On large scales (small $\ell$), the relative
errors are close to be unity, indicating that the Gaussian
approximation is reasonable. On larger $\ell$, however, the relative
errors increase and have peak at $k\sim 3-5h$/Mpc depending on the
halo mass.

We find that this excess of the covariance at small scales can be
explained by the halo sample variance (HSV). In the finite survey
area, mode fluctuations whose scales are larger than the survey area
are known to generate the excess covariance in the lensing auto power
spectra \citep{TakadaBridle07,Sato09}. The more complete formulae
called super sample covariance (SSC), which include the beat coupling
(BC) and the cross-term BC-HSV, was formulated in
\citet{TakadaHu13} and \citet{LiHuTakada14}. For the shear auto power
spectrum and galaxy-shear cross spectrum, HSV contributions to the
covariances are  written as \citep{TakadaHu13}
\begin{equation}
\label{eq:err_hsv1}
{\rm Cov}_{\kappa\kappa}^{\rm (HSV)}=
\int \frac{dz}{H(z)} \frac{W^{\kappa 4}(z)}{d_A^6(z)}
I_{\rm mm}(k_i,k_i)I_{\rm mm}(k_j,k_j)(\sigma_W^L(z))^2,
\end{equation}
and
\begin{eqnarray}
\label{eq:err_hsv2}
{\rm Cov}_{g\Sigma}^{\rm (HSV)}=
\int_{z_{\rm min}}^{z_{\rm max}} \frac{dz}{H(z)} [W^g(z)W^\Sigma(z)]^2 \frac{1}{d_A^6(z)} \nonumber \\
\times I_{gm}(k_i,k_i)I_{gm}(k_j,k_j)(\sigma_W^L(z))^2,
\end{eqnarray}
where $W^\Sigma (z)=\Sigma_{\rm crit}(z)W^\kappa$ and 
\begin{equation}
I_{\rm mm}(k,k')\equiv 
\int dM\frac{dn}{dM}\left(\frac{M}{\bar{\rho}_m}\right)^2 b(M,z) \tilde{u}_{\rm NFW}^2(k;M,z),
\end{equation}
and
\begin{eqnarray}
\label{eq:igm}
I_{gm}(k,k')\equiv \frac{1}{\bar{n}_g}
\int dM\frac{dn}{dM}\left(\frac{M}{\bar{\rho}_m}\right) (b(M,z)-1) \tilde{u}_{\rm NFW}(k;M,z) \nonumber \\
\left[\langle N_{\rm cen}\rangle + \langle N_{\rm sat}\rangle\tilde{p}_{\rm sat}(k';M,z)\right].
\end{eqnarray}
For the galaxy-shear cross power spectrum, we use $b(M)-1$ instead of
$b(M)$, which comes from local averaging of galaxy number
counts as discussed in the section II C in \citet{TakadaHu13}.
Since the local averaging rescales the observed power as
$P_W(k)=P(k)/(1+\delta_b)$, the HSV effect is reduced to some
extent. The response of the galaxy-galaxy lensing to the background is
modified as
\begin{equation}
\frac{\partial P(k)}{\partial \delta_b} \rightarrow 
\frac{\partial P(k)}{\partial \delta_b} - P(k), 
\end{equation}
which corresponds to converting $b(M)$ to $b(M)-1$.

The variance $(\sigma_W^L)^2$ represents the amplitude of 
the background fluctuation in a given survey window 
\begin{equation}
(\sigma_W^L(z))^2=\frac{1}{\Omega_{\rm sky}^2}\int_{k_{\rm min}}
\frac{\mathbf{dk}}{(2\pi)^2}|\tilde{W}(\mathbf{k})|^2 
P^L\left(k=\frac{\ell}{d_A(z)};z\right),
\end{equation}
where $P^{\rm L}(k;z)$ is the 3D linear power spectrum at $z$ and
$k_{\rm min}$ denotes the minimum wavenumber. The variance of the
background fluctuations $(\sigma_W^L)^2$ increases as the survey area
decreases. The ray-tracing simulations used in this paper are
constructed from $N$-body simulation boxes with the side length of 
$L_{\rm box}=240h^{-1}$Mpc up to $z=1$. In comparisons with the
simulation results, we set $k_{\rm min}=2\pi/L_{\rm box}$ with $L_{\rm
  box}=240h^{-1}$Mpc so that the fluctuations at scales larger than
$L_{\rm box}$ are excluded. As a result, $(\sigma_W^L)^2$ is decreased
roughly by half. When the field has a square shape, the survey window
function reads
\begin{equation}
\tilde{W}(\mathbf{k})=L^2{\rm sinc}(k_xL/2){\rm sinc}(k_yL/2),
\end{equation}
where $L$ is a side length of the square field. In this paper, we
approximate the square-shape survey window function with
$L=(\Omega_{\rm sky}^{\rm overlap})^{1/2}$.

In Fig.~\ref{fig:cl_sim_error}, we compare the ratio of the diagonal
elements of the covariance matrix from our simulations to the Gaussian
covariance with theoretical expectations of the enhancement of the
covariance matrix due to the HSV effect. Since we use dark haloes to
represent galaxies, in computing the HSV contributions the HOD
parameters are set to $\langle N_{\rm sat}\rangle=0$ for all $M_{\rm
  h}$ and $\langle N_{\rm cen}\rangle=1$ for $M_{\rm h}\ge
10^{13}h^{-1}M_\odot$ or $10^{14}h^{-1}M_\odot$ and $\langle N_{\rm
  cen}\rangle=0$ otherwise. We find that the simulations and HSV  
expectations agree reasonably well with each other, particularly when
$k_{\rm min}$ is set to the simulation box size. The HSV contribution
is comparable to the Gaussian term on large $l$ and thus the total
variance increases by 30-40\%. When the background fluctuation larger
than box size is included, which is relevant for the covariance in
real observations, the HSV contribution surpasses the Gaussian one and
the total variance doubles. 

\begin{figure*}
\begin{center}
\includegraphics[width=7cm]{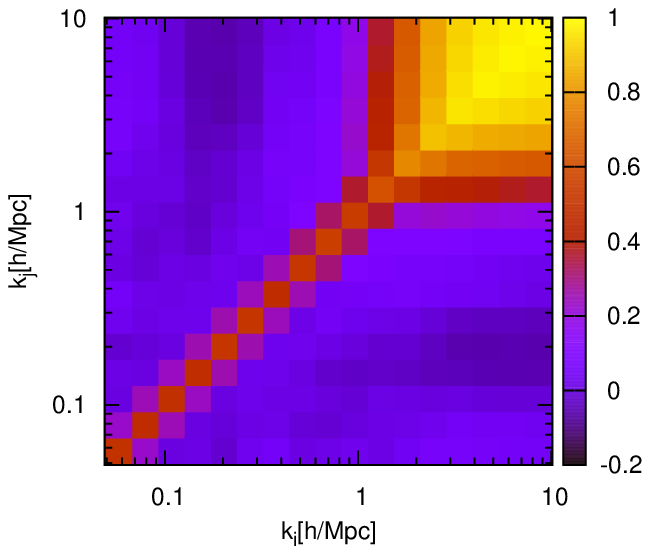}
\includegraphics[width=7cm]{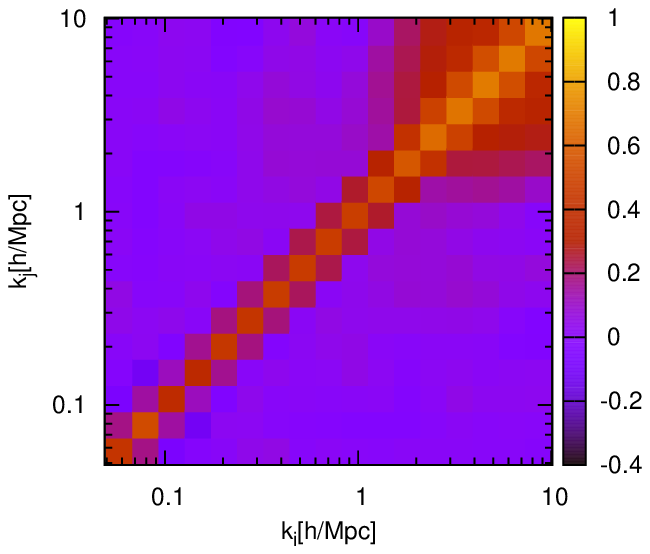}
\includegraphics[width=7cm]{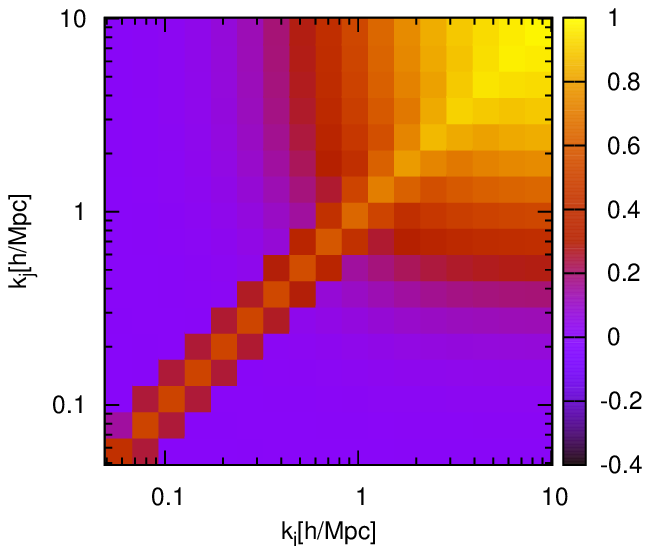}
\includegraphics[width=7cm]{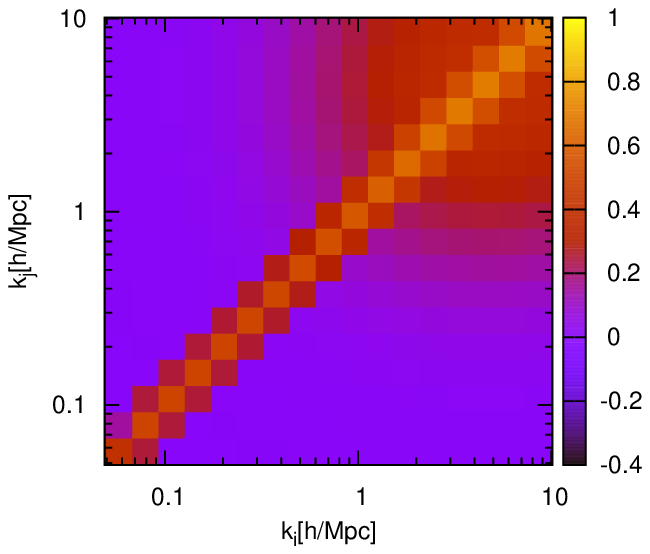}
\caption{Correlation coefficient matrix (eq.~[\ref{eq:coeff}]) of the
  galaxy-shear cross spectrum $P^{g\Sigma}(k)$ for $M_{\rm h}\ge
  10^{13}h^{-1}M_\odot$ with ({\it right panels}) and without shape noise
  ({\it left panels}). Upper panels show the correlation coefficient
  matrix derived from the simulated deconvolved galaxy-shear cross
  spectrum. Lower panels show the theoretical prediction of the
  covariance including the Gaussian covariance and the HSV
  contribution.} 
\label{fig:cl_cov}
\end{center}
\end{figure*}

More importantly, the HSV induces off-diagonal elements in the
covariance matrix, as shown in equation~(\ref{eq:err_hsv2}). We compute the
correlation coefficient matrix of the galaxy-shear cross spectrum as 
\begin{equation}
R_{ij}=\frac{{\rm Cov}_{g\Sigma}(k_i,k_j)}{
\sqrt{{\rm Cov}_{g\Sigma} (k_i,k_i){\rm Cov}_{g\Sigma}(k_j,k_j)}}.
\label{eq:coeff}
\end{equation}
Upper panels of Fig.~\ref{fig:cl_cov} show our simulation results on
 the covariance matrix for the galaxy-shear cross spectrum with the
halo mass $M_{\rm h}\ge 10^{13}h^{-1}M_\odot$. When the shape
noise is not included (left panel), the mode coupling due to the
nonlinear gravity increases the covariance among different scales at
larger $k$.  When the shape noise is included (right panel), the
covariance matrix approaches to be Gaussian. However, there are still
residuals in the off-diagonal components at small scales. We find that
the residual covariance is consistent with the predictions of Gaussian
plus the HSV term, as shown in the lower panels of Fig.~\ref{fig:cl_cov}.
This analysis suggests that the covariance matrix computed by adding
the Gaussian covariance and the HSV contribution provides a good
approximation to the covariance from our simulations which include
full non-Gaussian effects. We note that \citet{Gruen15} argued that 
intrinsic variations of the projected density profiles serve as
a source of off-diagonal covariance matrix at small scales. Our
analysis implies that the effect of the intrinsic variations is small
compared with the HSV contribution, at least in our setting where the
halo mass range is rather broad. In what follows, we use the analytic
formulae of the covariance Cov$^{\rm (G)}$+Cov$^{\rm HSV}$.  In the
following analysis, we show how the difference in the results with and
without the HSV contribution.

\section{Applications to CFHTLenS data and BOSS galaxy dataset}
\label{sec:cfhtlens}
\subsection{CFHTLenS data}
\label{subsec:CFHTLenS}
Here we use public available Canada-France-Hawaii Telescope Lensing
Survey (CFHTLenS) catalogue \citep{Erben13}. CFHTLenS data consists of
171 pointings (field of view of each pointing is $\sim 1$ deg$^2$) in
4 disjoint fields (W1,W2,W3,W4) covering an total effective survey
area of 154 deg$^2$. The imaging survey is carried out using 300 mega
pixels MEGACAM instrument with five filters and the $i$-band magnitude
limit $i_{AB}<24.5$ ($5\sigma$ in a $2''$ aperture).
 
We use the imaging data classified as a galaxy ({\it fitclass=0}) and
unmasked object ({\it MASK$\le$ 1}). The galaxy shape is measured
based on the {\it lensfit} algorithm \citep{Miller13} and the
ellipticity is defined as $e=(a-b)/(a+b)$ ($a$ and $b$ are major and
minor axes). The ellipticity data is obtained as a two-dimensional
vector $(e_1,e_2)$ where $e_1$ is the ellipticity component along the
constant declination and $e_2$ axis. Following the shear calibration
by \citet{Heymans12}, the additive term $c_2$ is subtracted from the
measured $e_2$.  Furthermore, each ellipticity component is divided by
the multiplicative term $m$ calibrated using simulations as a function
of image S/N and galaxy size \citep{Miller13}. The averaged values of
the galaxy shape ellipticity is 0.22 per component. We use the
ellipticity data with a positive value of the weight, which is
determined by the inverse of the variance due to the intrinsic galaxy
ellipticity and shape measurement error due to photon noise
\citep{Miller13}.

Since the galaxy-galaxy lensing measurement is much less affected by
the systematics than cosmic shear analysis, we use the data in the
whole of CFHTLenS fields. The photometric redshift of each galaxy is
estimated by the Bayesian Photometric Redshift (BPZ) code
\citep{Benitez00,Hildebrandt12}.  
%
We use source galaxies with photo-$z$ up to $z_{\rm BPZ}=3.1$ but the
probability $P_{\rm BPZ}(z>z_{\rm max})\ge 0.84$ where $z_{\rm max}$
is the maximum redshift of a given spectroscopic sample used as a
foreground lens sample for the galaxy-shear cross spectrum analysis
\citep[see also][for a similar background galaxy selection]{Oguri14}. 
The angular number density of source galaxies becomes $n_{\rm s}=7.2$
arcmin$^{-2}$ for CMASS and $9.6$ arcmin$^{-2}$ for LOWZ. 

We set a square field with a side length of 600 arcmin to cover each
CFHTLenS field.  The centre of the square field is defined as
$\alpha_c=(\alpha_{\rm max}+\alpha_{\rm min})/2$ and
$\delta_c=(\delta_{\rm max}+\delta_{\rm min})/2$ where $\alpha_{\rm
  max (min)}$ and $\delta_{\rm max (min)}$ is the maximum (minimum)
value of right ascension and declination of source galaxies in each
CFHTLenS field.  We convert the spherical coordinates to flat
coordinates as
$\cos(x)=\sin^2(\delta)+\cos^2(\delta)\cos(\alpha-\alpha_c)$ and
$y=\delta-\delta_c$ \citep{Kilbinger13}.

\subsection{Spectroscopic samples}
In order to cross-correlate with the CFHTLenS shear data, we use a
public catalogue of SDSS-III Data Release 11 BOSS spectroscopic galaxies
and the random catalogues \citep{Eisenstein11}.  We use two main
BOSS galaxy samples, ``LOWZ'' (lower redshift sample at $z<0.4$) and
``CMASS'' (higher redshift sample at $0.4<z<0.7$) \citep[see the
  details of the sample selection
  in][]{Eisenstein11,Ahn12,Dawson13}. We focus on the redshift range
of $0.16<z<0.33$ for LOWZ and $0.47<z<0.59$ for CMASS where the sample
is nearly volume-limited with a constant number density. We also use a
classical SDSS-I Luminous Red Galaxies (LRGs)
\citep{Eisenstein01}. Targeting cuts of the LRGs are similar to the
BOSS LOWZ and CMASS galaxies, though BOSS sample contains lower
stellar mass galaxies and their number density is about three times
higher than LRGs.

These spectroscopic samples partially overlap with the CFHTLenS
fields. The overlapped galaxy number becomes
418 for SDSS/LRG ($0.16<z<0.33$), 2353 for BOSS/LOWZ ($0.16<z<0.33$), 
and 5429 for BOSS/CMASS ($0.47<z<0.59$) samples. 
The overlapped sky fraction of CFHTLenS field is 35 \% for LRG, 74\%
for LOWZ and 77\% for CMASS sample. There is almost no overlap for the
field W2 and thus we do not use W2 field in our galaxy-galaxy lensing
analysis.

\begin{figure}
\begin{center}
\includegraphics[width=8cm]{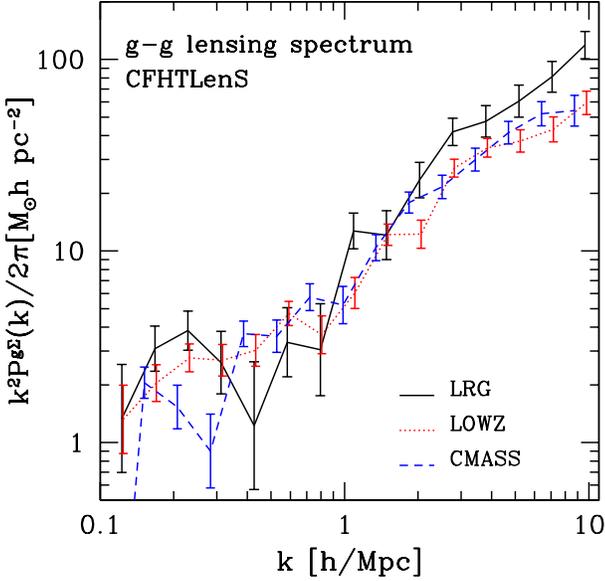}
\caption{The galaxy-galaxy lensing pseudo spectrum, i.e., the galaxy-shear
  cross spectrum with the correction of the masking effect, between
  the CFHTLenS shear catalogue and three different spectroscopic galaxy
  samples, SDSS DR7 LRG ({\it solid}), BOSS/LOWZ ({\it dotted}),  and
  BOSS/CMASS ({\it dashed}).}
\label{fig:cl_hg_obsonly}
\end{center}
\end{figure}

\subsection{Results}
We measure the galaxy-shear cross spectrum $P^{g\Sigma}(k)$, i.e., the
galaxy-galaxy lensing profile in the Fourier space, using the
pseudo-spectrum method. Fig.~\ref{fig:cl_hg_obsonly} show the
comparison of measured cross spectra between the CFHTLenS shear
catalogue and three spectroscopic samples: BOSS/CMASS ($0.47<z<0.59$),
BOSS/LOWZ ($0.16<z<0.33$), and SDSS DR7 LRG ($0.16<z<0.33$). When
converting from $C_\ell^{g\kappa}$ to $P^{g\Sigma}$, for simplicity we
use the mean redshift $\bar{z}$ of each galaxy sample assuming a flat
$\Lambda$CDM cosmology with $\Omega_m=0.3$ (eq.~[\ref{eq:pk_gsig}]).
We find that the cross-correlation with the SDSS LRG sample has larger
amplitude at larger $k$. This is consistent with that the SDSS/LRGs
are hosted by more massive haloes than the other two samples
\citep[see also][]{Miyatake15}.

We estimate the signal-to-noise ratio for the galaxy-shear cross
spectrum as
\begin{equation}
({\rm S/N})^2=\sum_{i,j}P^{g\Sigma}(k_i)({\rm Cov^{g\Sigma}}^{-1})_{ij}P^{g\Sigma}(k_j).
\end{equation}
We use the range of scales from $0.1h^{-1}$Mpc to 10$h^{-1}$Mpc with
15 data points. Here the covariance is estimated analytically using
the Gaussian covariance and the HSV contribution (see
Section~\ref{sec:cov}). The S/N of the galaxy-galaxy lensing for the
three galaxy samples are 7.1 (SDSS/LRG), 8.6 (BOSS/LOWZ), and 10.4
(BOSS/CMASS). For comparison, the S/N increases to 18.1 (SDSS/LRG),
23.7 (BOSS/LOWZ), 20.4 (BOSS/CMASS) without the HSV term in the
covariance. The S/N value decreases nearly by half when including the HSV,
which indicate that the effect of super survey modes is important also
for the galaxy-galaxy lensing analysis.  Even when the HSV is taken
into account, we find that galaxy-galaxy lensing signals are detected
at the significance level of $7-10\sigma$ for all of the three
spectroscopic samples. 

We compare the measured spectra with the halo model calculations
assuming a commonly used HOD form with 5 parameters by \citet{Zheng05} 
\begin{eqnarray}
\label{eq:hod}
\langle N_{\rm cen}\rangle &=&\frac{1}{2}\left[1+{\rm erf}\left(\frac{\log_{10}(M)-\log_{10}
(M_{\rm c})}{\sigma_{\log M}}\right)\right], \nonumber \\
\langle N_{\rm sat}\rangle &=&\langle N_{\rm cen}\rangle\left(\frac{M-M_{\rm 0}}{M_1}\right)^{\alpha},
\end{eqnarray}
where ${\rm erf}(x)$ is the error function and their HOD parameters in
the previous work are listed in Table~\ref{tab:hod}.  The HOD
parameter values for LRGs \citep{Reid09a} is estimated using the
counts-in-cylinder group finding technique based on {\it WMAP3}
cosmology, while the others are estimated from redshift-space
clustering based on {\it WMAP7} cosmology \citep{Manera13,Manera15}.
When comparing the HOD model predictions with our measurements, we use
the same cosmological parameters as those assumed when deriving the
HOD parameter values.  The difference of the cosmology is included in
the conversion from $C_\ell^{g\kappa}$ to $P^{g\Sigma}(k)$ as
$P^{g\Sigma}(k)\rightarrow P^{g\Sigma}(k)[d_A^2\Sigma_{\rm crit}]^{\rm
  (fid)}/[d_A^2\Sigma_{\rm crit}]$ and $k\rightarrow kd_A(z)/d_A^{\rm
  (fid)}(z)$, though the conversion effect is much smaller than the
statistical error.

\begin{table*}
\begin{center}
\begin{tabular}{cccc}
\hline
\hline
parameters & SDSS/LRG & BOSS/LOWZ & BOSS/CMASS \\
 & \citep{Reid09a} & \citep{Manera15} & \citep{Manera13} \\
\hline
cosmology & {\it WMAP3} & {\it WMAP7} & {\it WMAP7} \\
$\log_{10}(M_{\rm c})$  & 13.76 & 13.2 & 13.09  \\
$\sigma_{\log M}$  & 0.7 & 0.62  & 0.60   \\
$\log_{10}(M_{\rm 0})$  & 13.54 & 13.24 & 13.08  \\
$\log_{10}(M_1)$        & 14.54 & 14.32 & 14.00  \\
$\alpha$ & 1.03  & 0.93  & 1.01   \\
\hline
$\chi^2$(d.o.f.=15) & 14.2 & 20.7 & 17.2 \\
\hline
\end{tabular}
\caption{The reference values of the 5 HOD parameters for the three
  spectroscopic samples, SDSS/LRGs \citep{Reid09a}, BOSS/CMASS, and
  LOWZ \citep{Manera13,Manera15}. The cosmology based on {\it WMAP}
  3-year is assumed for the HOD parameter values of SDSS/LRGs, while the other
  HOD parameter values are based on {\it WMAP} 7-year
  \citep{Komatsu11}. The unit of mass is $h^{-1}M_\odot$. The
  $\chi^2$ values are obtained by comparing the HOD model predictions
  with the observed galaxy-shear cross spectra from $0.1h$/Mpc to
  $10h$/Mpc (15 bins of $k$).} 
\label{tab:hod}
\end{center}
\end{table*}

\begin{figure*}
\begin{center}
\includegraphics[width=7cm]{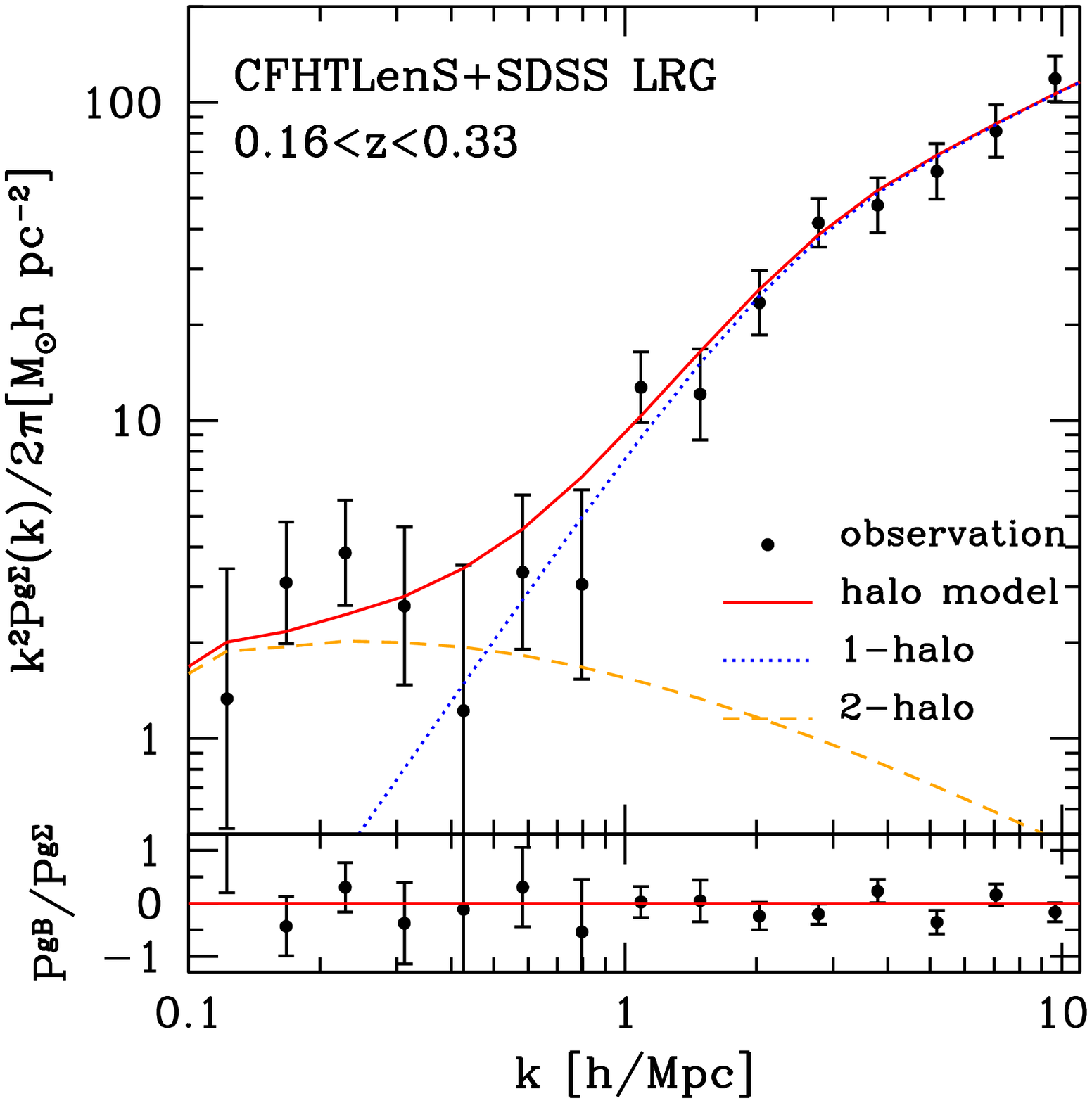}
\includegraphics[width=7cm]{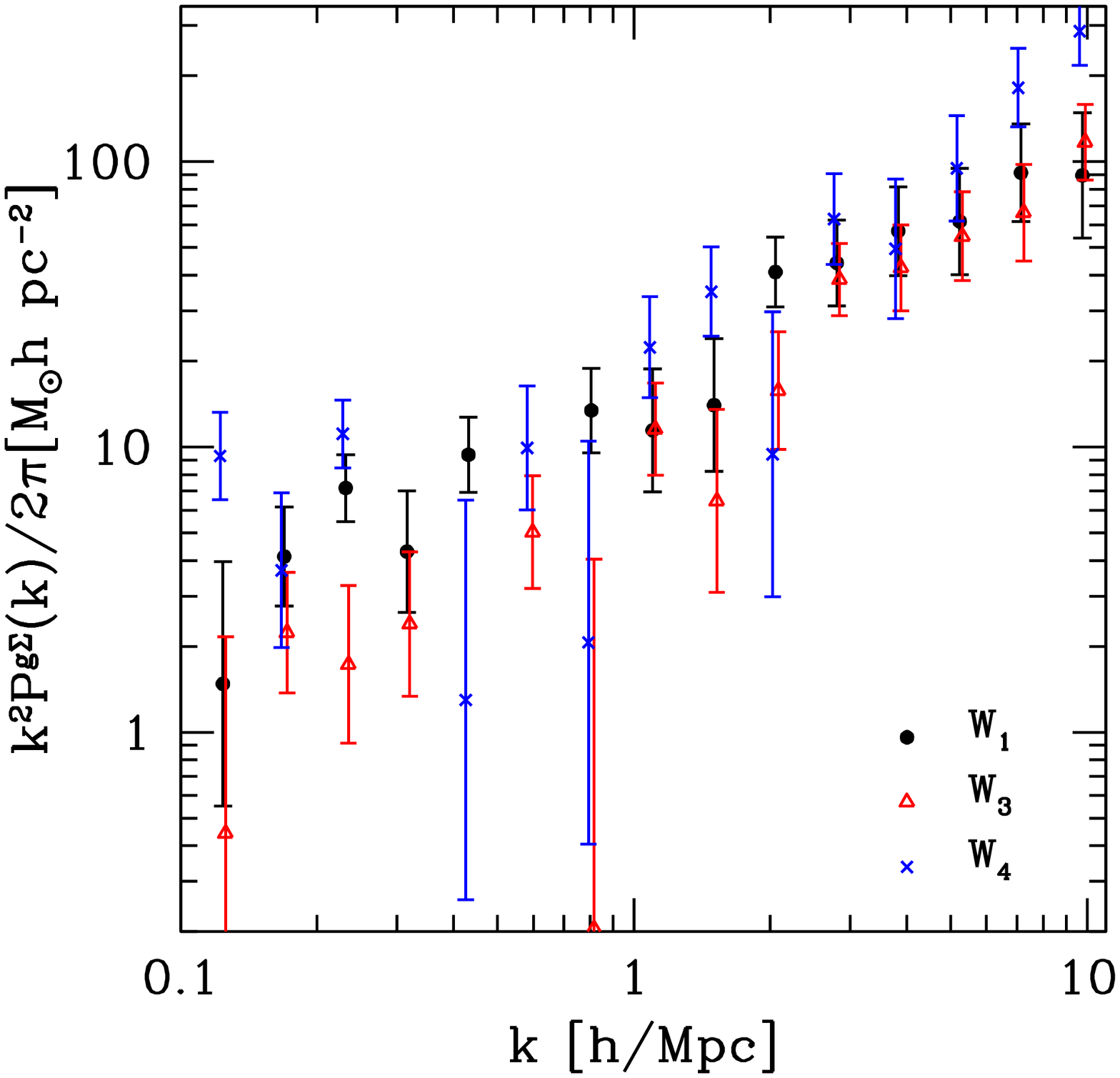}
\includegraphics[width=7cm]{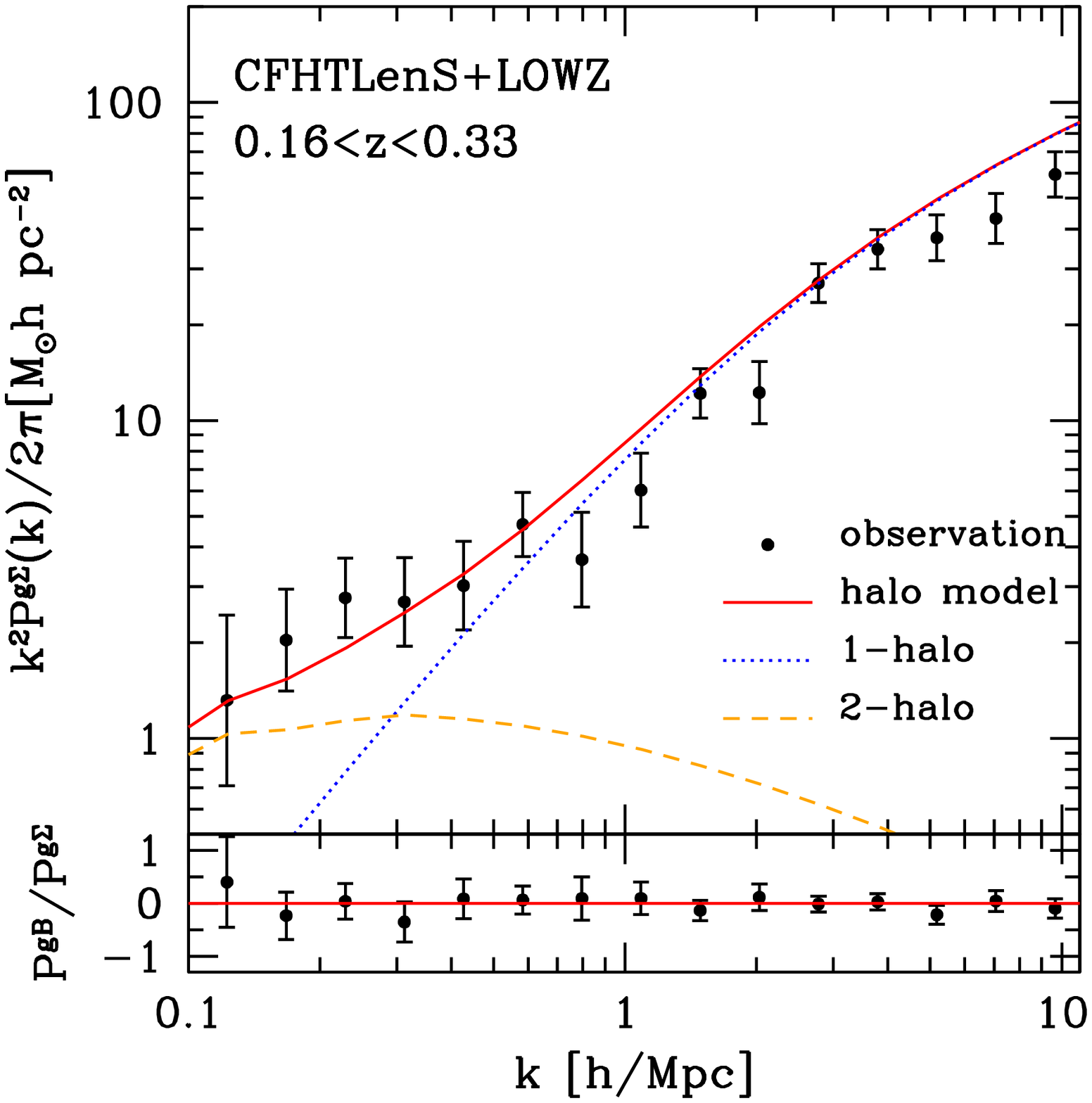}
\includegraphics[width=7cm]{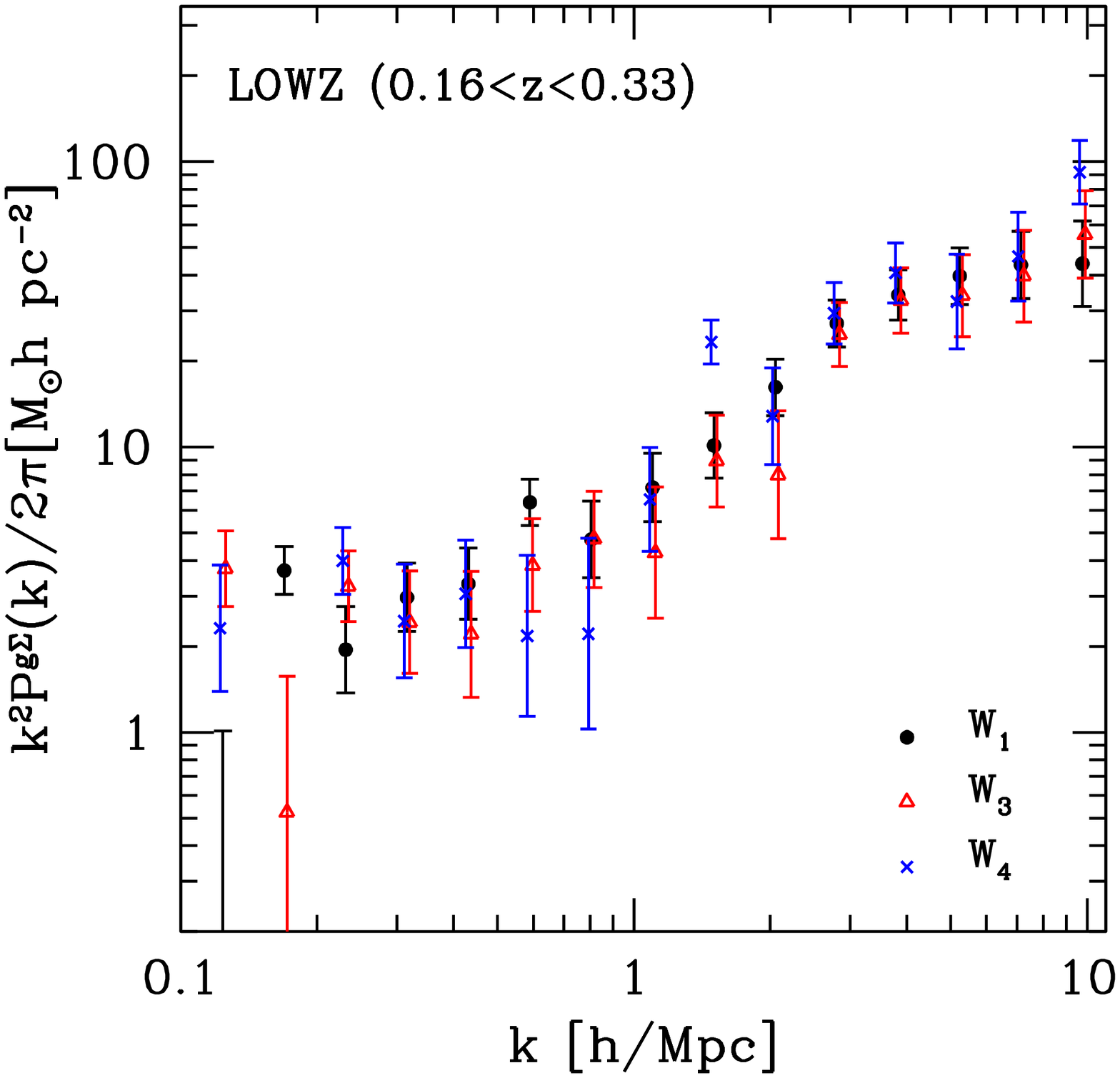}
\includegraphics[width=7cm]{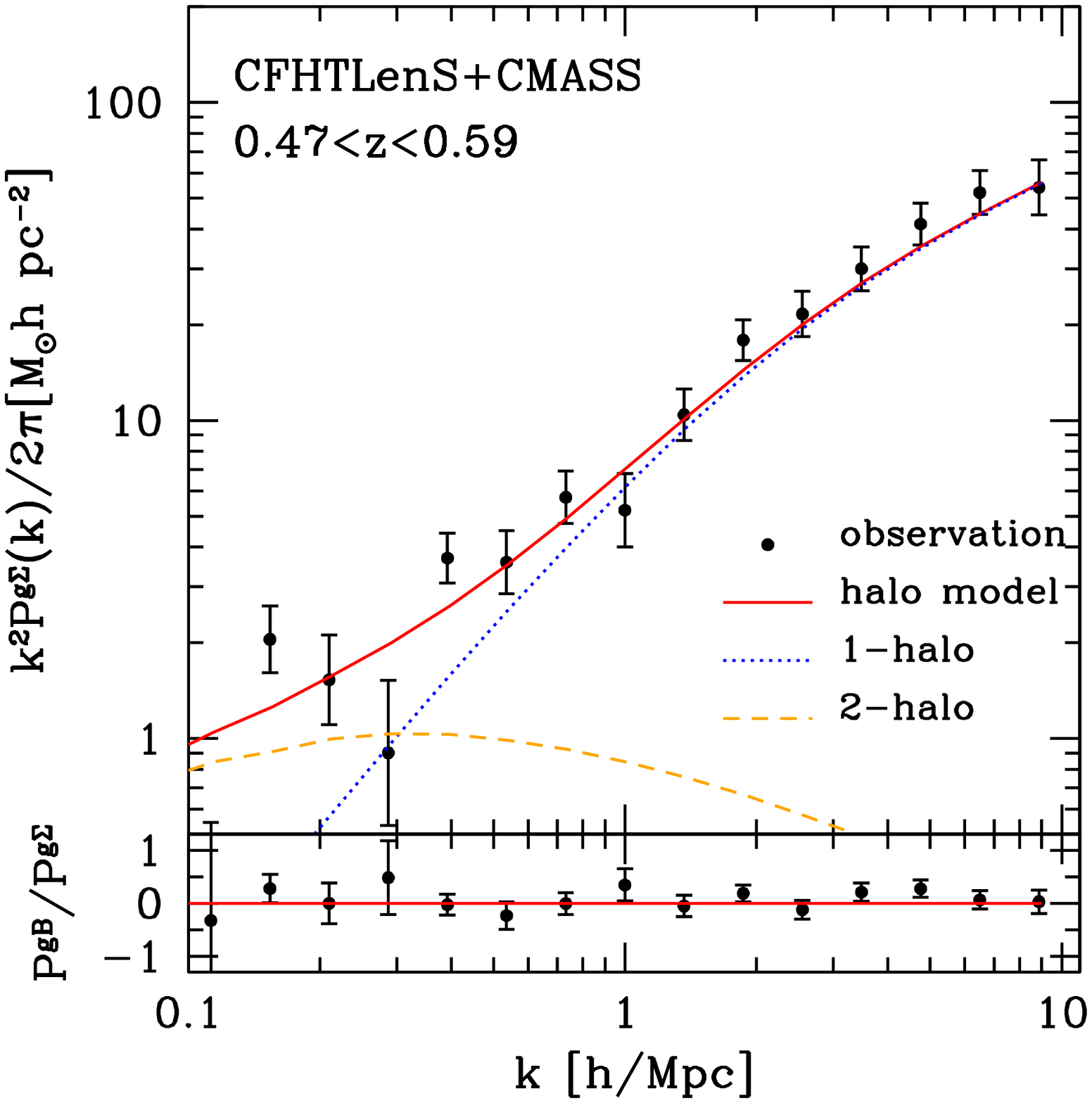}
\includegraphics[width=7cm]{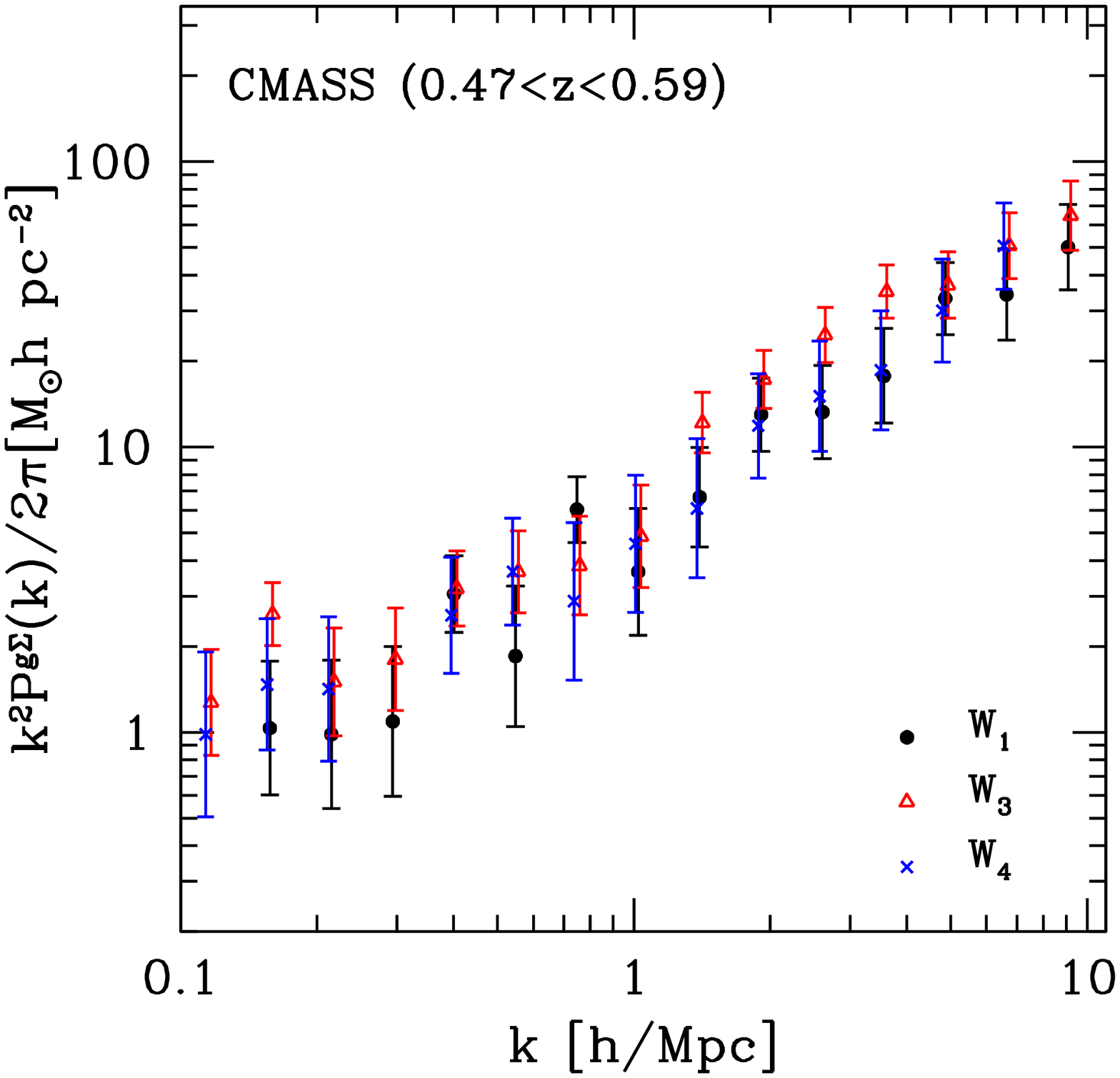}
\caption{{\it Left:} Comparisons of the observed galaxy-shear cross spectra
  with the HOD model. From top to bottom, we show the results for
  SDSS/LRG, BOSS/LOWZ, and BOSS/CMASS. The HOD model parameters used
  here are summarized in Table~\ref{tab:hod}. The one-halo term,
  two-halo term, and their sum are plotted with blue dotted, yellow
  dashed, and red solid lines, respectively. The observed cross
  spectra between the B-mode shear and galaxy density fields are also
  shown at the bottom panel of each figure. {\it Right:} The
  difference of the observed galaxy-shear cross spectra between three
  CFHTLenS fields (W1,W3,W4), again for the same three galaxy samples.}
\label{fig:cl_hg_obs}
\end{center}
\end{figure*}

Fig.~\ref{fig:cl_hg_obs} compares the HOD model predictions with
observed spectra.  Individual contributions from the one-halo and
two-halo components are plotted separately. We find that the halo
model with the HOD parameters in previous work well explains our
measured galaxy-shear cross spectra. To quantify the goodness, we
estimate the $\chi^2$ value as 
\begin{eqnarray}
\label{eq:chi2}
\chi^2=\sum_{i,j}\Delta P^{g\Sigma}(k_i)({\rm Cov}^{g\Sigma~-1})_{ij}\Delta P^{g\Sigma}(k_j),
\end{eqnarray}
where $\Delta P^{g\Sigma}(k)$ is defined by the difference between the
observed spectrum and the model spectrum.  For all of the three
spectroscopic galaxy samples, we obtain reasonable $\chi^2$ values as
listed in Table~\ref{tab:hod}. The consistency of the HOD between the
different measurements indicates that the halo model description works
well in the current uncertainty level. This consistency can also serve
as a sanity check that our pseudo-spectrum method works well with real
observational data. The right panels in Fig.~\ref{fig:cl_hg_obs} show
that the comparison of galaxy-shear cross spectra for three CFHTLenS
fields (W1,W3,W4). We find that the galaxy-shear cross spectra are
detected in each field and they are consistent with each other.

\section{Conclusion}
\label{sec:conclusion}
In this paper, we have presented the formalism of galaxy-galaxy
lensing in the Fourier space. Our approach takes account of the
realistic mask effect using the pseudo-spectrum method.  Using
ray-tracing simulations in $N$-body simulations, we have confirmed
that the pseudo-spectrum method successfully recovers the input true
galaxy-shear cross spectrum, which is a Fourier space counterpart of
the stacked galaxy-galaxy lensing profile. We note that our formalism
allows different mask patterns between the shear and galaxy number
density fields.

We have also investigated the covariance of the galaxy-shear cross
spectrum using the ray-tracing simulations. We have found the excess
covariance relative to the Gaussian error on large $k$ where the shot
noise is dominated in the Gaussian approximation. We have shown that
the excess can be explained by the halo sample variance (HSV), which
originates from matter fluctuations at scales larger than the survey
area. In our examples, the HSV contribution increases the diagonal
error of galaxy-galaxy lensing nearly twice at $k\la 1h$/Mpc, and also
induces the off-diagonal elements.  The HSV contribution has been
ignored in previous galaxy-galaxy lensing analysis, and our results
highlight the importance of including super survey modes in the
covariance. We note that the Jackknife method does not contain
information on fluctuations larger than the survey area, and thus
substantially underestimates the HSV contribution.  While we have
included only the HSV term, more comprehensive analysis of the effect
of the super sample covariance on galaxy-galaxy lensing is ongoing
(Takada et al. in preparation).

We have applied the pseudo-spectrum method to the real observational
dataset from the CFHTLenS shear catalogue and various spectroscopic
samples including SDSS/LRG, BOSS/LOWZ, and BOSS/CMASS. We have
detected the galaxy-shear cross spectra for all the three spectroscopic
samples at the significance level of $7-10\sigma$ by using the
analytic covariance formulae including the HSV effect. 
We have confirmed that our galaxy-galaxy lensing measurements are
consistent with the theoretical predictions based on the HOD model
from previous work, which can be seen as a sanity check of our
pseudo-spectrum method to measure the galaxy-galaxy lensing in the Fourier
space. The methodology developed in this paper will be important for
analyzing future data with which we will be able to measure
galaxy-shear cross spectra out to larger scales. 

\section*{Acknowledgments}
We thank Masahiro Takada for fruitful discussions. 
This work was supported in part by World Premier International
Research Center Initiative (WPI Initiative), MEXT, Japan, and
JSPS KAKENHI Grant Number 26800093 and 15H05892.

\bibliographystyle{mn2e} \bibliography{mn-jour,ref} \label{lastpage}

\appendix
\section{Full-sky formalism} 
\label{sec:app}
In this Appendix, we provide the full-sky formalism for the
pseudo-spectrum analysis of the shear power spectrum and galaxy-shear
cross spectrum. This is analogous to the formalism developed for the
CMB polarization and temperature cross spectrum
\citep{Hivon02,Kogut03,Brown05}.

The observed shear field $\Bgamma$ can be decomposed into E-mode and
B-mode shear by spherical harmonic transform with the spin-2 spherical
harmonics ${}_{\pm 2}Y_{\ell m}$
\begin{equation}
\tilde{E}_{\ell m}\pm i\tilde{B}_{\ell m}=\int d\hat\mathbf{n}~
\Bgamma(\hat\mathbf{n})~{}_{\pm 2}Y_{\ell m}^\ast(\hat\mathbf{n}),
\end{equation}
and the inverse relation is
\begin{equation}
\Bgamma(\hat\mathbf{n})=\sum_{\ell m}(\tilde{E}_{\ell m}\pm i\tilde{B}_{\ell m}){}_{\pm 2}Y_{\ell m}(\hat\mathbf{n}).
\end{equation}
The relation between the galaxy number density field and their
Harmonic coefficients are given using the spin-0 spherical harmonics ${}_0Y_{\ell m}$ as
\begin{equation}
\tilde{\delta}_{g,\ell m}=\int d\hat\mathbf{n}~\delta_g(\hat\mathbf{n})~{}_0Y_{\ell m}^\ast(\hat\mathbf{n}),
\end{equation}
and
\begin{equation}
\delta_g(\hat\mathbf{n})=\sum_{\ell m} \tilde\delta_{g,\ell m}~{}_0Y_{\ell m}(\hat\mathbf{n}).
\end{equation}
The effect of survey mask on the harmonic coefficients is expressed as their convolution
\begin{eqnarray}
(\tilde{E}_{\ell m}\pm i\tilde{B}_{\ell m})^{\rm (obs)}=\int d\hat\mathbf{n}
U^\gamma(\hat\mathbf{n})\Bgamma(\hat\mathbf{n}) {}_{\pm 2}Y_{\ell m}^\ast(\hat\mathbf{n}) \\
=\sum_{\ell'm'}(E_{\ell'm'}\pm iB_{\ell'm'})^{\rm (true)} {}_{\pm 2}W_{\ell\ell'mm'}^\gamma 
\end{eqnarray}
and
\begin{eqnarray}
\tilde\delta_g^{\rm (obs)}(\hat\mathbf{n})=\int d\hat\mathbf{n}~
U^g(\hat\mathbf{n})\delta_g(\hat\mathbf{n}) {}_0Y_{\ell m}^\ast(\hat\mathbf{n}) \\
=\sum_{\ell'm'}\tilde\delta_{g,\ell'm'}^{\rm (true)}){}_{0}W_{\ell\ell'mm'}^g,~~~~~~~~~~~~
\end{eqnarray}
and the convolution kernels are defined as 
\begin{eqnarray}
{}_sW_{\ell\ell'mm'}^{g(\gamma)}\equiv \int d\hat\mathbf{n}~
{}_sY_{\ell' m'}(\hat\mathbf{n})
U^{g(\gamma)}(\hat\mathbf{n})
{}_sY_{\ell m}^\ast(\hat\mathbf{n})~~~~~~~~~~~~~~ \nonumber \\
=\sum_{\ell''m''}\tilde{U}_{\ell''m''}^{g(\gamma)}(-1)^{m}
\sqrt{\frac{(2\ell+1)(2\ell'+1)(2\ell''+1)}{4\pi}} \nonumber \\
\times 
\left(\begin{array}{ccc}
\ell & \ell' & \ell'' \\
s & -s & 0
\end{array}\right)
\left(\begin{array}{ccc}
\ell & \ell' & \ell'' \\
m & m' & m''
\end{array}\right),
\end{eqnarray}
where $\tilde{U}^{g(\gamma)}_{\ell m}$ represents the spherical harmonic transform of 
the mask field for galaxy overdensity field $g$ or shear field $\gamma$ 
\begin{equation}
\tilde{U}^{g(\gamma)}_{\ell m}=\int d\hat\mathbf{n}~U^{g(\gamma)}(\hat\mathbf{n}){}_0Y_{\ell m}^\ast(\hat\mathbf{n}).
\end{equation}
The auto and cross spectra are defined as
\begin{equation}
C_\ell^{XY}\equiv \frac{1}{2\ell+1}\sum_m \ave{X_{\ell m}Y_{\ell m}^\ast},
\end{equation}
where $X$ and $Y$ denotes $E$-mode shear, $B$-mode shear, and galaxy overdensity $g$. The pseudo spectra computed from 
the observed masked field are related to the true spectra as
\begin{equation}
\mathbf{C}_\ell^{\rm (obs)}=\sum_{\ell'}\mathbf{M}_{\ell\ell'}F_{\ell'}^2\mathbf{C}_{\ell'}^{\rm (true)}+\mathbf{N}_\ell^{\rm (obs)}.
\end{equation}
where $\mathbf{C}$ denotes the 6-dimensional vector of power spectra
for E-mode shear, B-mode shear, and galaxy overdensity and their cross
spectra.  Non-zero components of the mode coupling matrix
$\mathbf{M}_{\ell\ell'}$ are as follows:
\begin{eqnarray}
M_{\ell\ell'}^{EE,EE}&=&M_{\ell\ell'}^{BB,BB} \qquad  \nonumber \\
&=& \frac{2\ell'+1}{8\pi}\sum_{\ell''}{\cal U}_{\ell''}^{\gamma\gamma}[1+(-1)^{\ell+\ell'+\ell''}]
\left(\begin{array}{ccc}
\ell & \ell' & \ell'' \\
2 & -2 & 0
\end{array}\right)^2, \nonumber \\ \\
M_{\ell\ell'}^{EE,BB}&=&M_{\ell\ell'}^{BB,EE} \nonumber \\
&=& \frac{2\ell'+1}{8\pi}\sum_{\ell''}{\cal U}_{\ell''}^{\gamma\gamma}[1-(-1)^{\ell+\ell'+\ell''}]
\left(\begin{array}{ccc}
\ell & \ell' & \ell'' \\
2 & -2 & 0
\end{array}\right)^2, \nonumber \\ \\
M_{\ell\ell'}^{EB,EB}&=&\frac{2\ell'+1}{4\pi}\sum_{\ell''}{\cal U}_{\ell''}^{\gamma\gamma}
\left(\begin{array}{ccc}
\ell & \ell' & \ell'' \\
2 & -2 & 0
\end{array}\right)^2, \\
M_{\ell\ell'}^{gE,gE}&=&M_{\ell\ell'}^{gB,gB} \nonumber \\
&=&\frac{2\ell'+1}{4\pi}\sum_{\ell''}{\cal U}_{\ell''}^{g\gamma}
\left(\begin{array}{ccc}
\ell & \ell' & \ell'' \\
0 & 0 & 0
\end{array}\right)^2 
\left(\begin{array}{ccc}
\ell & \ell' & \ell'' \\
2 & -2 & 0
\end{array}\right)^2, \nonumber \\ \\
M_{\ell\ell'}^{gg,gg}&=&\frac{2\ell'+1}{4\pi}\sum_{\ell''}{\cal U}_{\ell''}^{gg}
\left(\begin{array}{ccc}
\ell & \ell' & \ell'' \\
0 & 0 & 0
\end{array}\right)^2.
\end{eqnarray}
where
\begin{equation}
{\cal U}_{\ell}^{XY}\equiv \frac{1}{2\ell+1}\sum_m\langle \tilde{U}_{\ell m}^X \tilde{U}_{\ell m}^{\ast Y}\rangle,
\end{equation}
with $X$ and $Y$ being $g$ or $\gamma$.

\end{document}